\documentclass[fleqn,usenatbib]{mnras}

\usepackage{newtxtext,newtxmath}

\usepackage[T1]{fontenc}

\DeclareRobustCommand{\VAN}[3]{#2}
\let\VANthebibliography\thebibliography
\def\thebibliography{\DeclareRobustCommand{\VAN}[3]{##3}\VANthebibliography}

\usepackage{siunitx}
\usepackage{graphicx}	
\usepackage{amsmath}	
\usepackage{arydshln}          
\usepackage[version=4]{mhchem}
\usepackage[capitalise]{cleveref}

\title[Transport-induced chemistry signatures in 3D]{Observability of signatures of transport-induced chemistry in clear atmospheres of hot gas giant exoplanets}

\author[Zamyatina et al.]{Maria Zamyatina,$^{1}$\thanks{E-mail: m.zamyatina@exeter.ac.uk}
Eric H\'ebrard,$^{1}$
Benjamin Drummond,$^{2}$
Nathan J. Mayne,$^{1}$
James Manners,$^{2}$
\newauthor
Duncan A. Christie,$^{1}$
Pascal Tremblin,$^{3}$
David K. Sing$^{4,5}$
and Krisztian Kohary$^{1}$
\\
$^{1}$Department of Physics and Astronomy, Faculty of Environment, Science and Economy, University of Exeter, Exeter EX4 4QL, UK\\
$^{2}$Met Office, Fitzroy Road, Exeter, EX1 3PB, UK\\
$^{3}$Maison de la Simulation, CEA, CNRS, Univ. Paris-Sud, UVSQ, Université Paris-Saclay, F-91191 Gif-sur-Yvette, France\\
$^{4}$Department of Earth and Planetary Science, Johns Hopkins University, 3400 N. Charles Street, Baltimore, MD 21218, USA\\
$^{5}$Department of Physics and Astronomy, Johns Hopkins University, 3400 N. Charles Street, Baltimore, MD 21218, USA
}

\date{Accepted XXX. Received YYY; in original form ZZZ}

\pubyear{2022}

\begin{document}
\label{firstpage}
\pagerange{\pageref{firstpage}--\pageref{lastpage}}
\maketitle

\begin{abstract}
Transport-induced quenching, i.e., the homogenisation of chemical abundances by atmospheric advection, is thought to occur in the atmospheres of hot gas giant exoplanets. While some numerical modelling of this process exists, the three-dimensional nature of transport-induced chemistry is underexplored. Here we present results of 3D cloud- and haze-free simulations of the atmospheres of HAT-P-11b, HD~189733b, HD~209458b, and WASP-17b including coupled hydrodynamics, radiative transfer and chemistry. Our simulations were performed with two chemical schemes: a chemical kinetics scheme, which is capable of capturing transport-induced quenching, and a simpler, more widely used chemical equilibrium scheme. We find that transport-induced quenching is predicted to occur in atmospheres of all planets in our sample; however, the extent to which it affects their synthetic spectra and phase curves varies from planet to planet. This implies that there is a ``sweet spot'' for the observability of signatures of transport-induced quenching, which is controlled by the interplay between the dynamics and chemistry.
\end{abstract}




\section{Introduction}

Atmospheric characterisation is one of the current top-level challenges of exoplanet science \citep{Howell2020}. It calls for a coordinated effort by both sides of the exoplanet community, observational and theoretical, to obtain a better understanding of the exoplanets' optical structure, measured with transmission spectroscopy, emission and reflection spectroscopy, and phase curves \citep{Kreidberg2018} and simulated with a hierarchy of numerical models.

It is becoming apparent that to interpret observations of the best observational targets, currently Jovian planets in short-period orbits, or hot Jupiters, our community requires multidimensional models that couple hydrodynamics, radiative transfer, and chemistry. Evidence of such a coupling having an impact on the observed properties of hot Jupiter atmospheres had been obtained. For example, vertical mixing was shown to cause a departure from chemical equilibrium in the atmosphere of Jupiter \citep{Prinn1977} and hot Jupiter analogues \citep[e.g.,][]{Saumon2003,Skemer2014,Zahnle2014}. Horizontal mixing, specifically prograde equatorial winds, was proven to occur in atmospheres of hot Jupiters \citep[e.g.,][]{Snellen2010,Wyttenbach2015,Louden2015}, and reported to lead to departures from chemical equilibrium with 2D \citep[e.g.,][]{Agundez2012,Agundez2014,Baeyens2021,Baeyens2022} and 3D forward models \citep[e.g.,][]{Cooper2006,Drummond2018b,Drummond2018a,Mendonca2018,Steinrueck2018,Drummond2020}. 1D forward models and 1D retrievals still surpass higher dimensionality modelling in terms of computational efficiency, but do that at a cost of insufficiently capturing the 3D structure of an observed atmosphere \citep{Pluriel2020,MacDonald2020,Irwin2020}.

Recently, \citet{Drummond2020} presented 3D cloud- and haze-free coupled hydrodynamics-radiation-chemistry simulations of atmospheres of HD~189733b and HD~209458b. They showed that the choice of a chemical scheme, between the chemical equilibrium scheme (allowing pressure and temperature to affect the chemical structure) and the chemical kinetics scheme (allowing pressure, temperature and 3D mixing to impact the chemical structure), resulted in synthetic observations from telescopes such as JWST \citep{Greene2016} and Ariel \citep{Tinetti2018,Charnay2022} to have potentially detectable differences between their respective simulations. Although more work is required to constrain, improve and learn from such simulations, the results for the somewhat similar, in terms of bulk planetary parameters, HD~189733b and HD~209458b simulations showed a surprising level of sensitivity in the observable chemical composition.

In this study, we follow up on this finding, the apparent planet-dependency of the observability of signatures of transport-induced chemistry, and simulate the atmospheres of two additional exoplanets with the same model and approach as in \cite{Drummond2020}. To expand the explored parameter space, we choose a warm super-Neptune HAT-P-11b discovered by \cite{Bakos2010} as the smallest and coldest exoplanet in our sample of four exoplanets, and a hot Jupiter WASP-17b discovered by \cite{Anderson2010} as the largest and hottest exoplanet.

The outline of this paper is as follows. \cref{section:methods} introduces our model and details the parameters of our four targets. \cref{section:results} describes the simulated HAT-P-11b, HD~189733b, HD~209458b, and WASP-17b atmospheres and compares their dynamical (\cref{section:dynamical_structure}), thermal (\cref{section:thermal_structure}), and chemical structure (\cref{section:chemical structure}), as well as their observed and synthetic transmission spectra (\cref{section:transmission}), emission spectra (\cref{section:emission}), and phase curves (\cref{section:phase_curves}). \cref{section:discussion} discusses the observability of signatures of transport-induced quenching. \cref{section:conclusions} states our conclusions and plans for future work.

\section{Methods}
\label{section:methods}

\subsection{Model description}
\label{section:methods_model_description}

We employ an idealised configuration of the Met Office Unified Model (UM), whose dynamical \citep[ENDGame,][]{Wood2014,Mayne2014,Mayne2014a,Mayne2017,Mayne2019}, radiative \citep[SOCRATES,][]{Edwards1996,Edwards1996a,Amundsen2014,Amundsen2016,Amundsen2017} and chemical \citep{Drummond2018,Drummond2018a,Drummond2018b,Drummond2020} components have been adapted, tested and applied to study gas giant atmospheres. The dynamical core of the model uses a semi-implicit semi-Lagrangian scheme to solve the full, deep-atmosphere, non-hydrostatic equations of motion discretised horizontally onto a regular longitude-latitude Arakawa-C staggered grid and vertically onto a geometric height-based Charney-Phillips staggered grid. At the poles only the meridional component of the wind is stored, and its speed and direction are determined by a least-squares best fit to the zonal wind on the grid-row closest to the pole assuming that the wind across the pole is that of a solid-body rotation. A diffusion scheme suppresses grid-scale numerical instabilities in horizontal wind velocities in the zonal direction only, with minimal additional damping applied to vertical wind velocities near the upper boundary and no damping or frictional parameterisation at the bottom boundary. The radiative transfer component of the model solves the two-stream equations, considers the absorption due to \ce{H2O}, \ce{CO}, \ce{CO2}, \ce{CH4}, \ce{NH3}, \ce{HCN}, \ce{Li}, \ce{Na}, \ce{K}, \ce{Rb}, \ce{Cs} and collision-induced absorption due to \ce{H2}-\ce{H2} and \ce{H2}-\ce{He} using the correlated-k and equivalent extinction methods for computing k-coefficients from the ExoMol line lists \citep{Tennyson2016} (see \cref{Appendix_A} for details), and includes Rayleigh scattering due to \ce{H2} and \ce{He}. Alkali metal abundances are computed from their respective monatomic/polyatomic transition boundaries using the \citet{Burrows1999} analytical fits, with additional smoothing applied to prevent non-continuous abundance change. We do not include \ce{TiO} and \ce{VO} as opacity sources because our targets' planetary effective temperatures are low enough ($<$\SI{1700}{\kelvin}) to allow us to assume \ce{TiO} and \ce{VO} to be rained out \citep{Burrows1999,Goyal2019}.

The chemical component of the model provides a selection of chemical schemes, from the most simple to the more complete, but computationally demanding: (1) analytical, (2) equilibrium, (3) relaxation, and (4) kinetics. The \textit{analytical} chemical scheme uses the \citet{Burrows1999} chemical equilibrium abundance formulas for \ce{H2O}, \ce{CO}, \ce{CH4} and \ce{NH3} \citep[see, e.g.,][]{Amundsen2016}. This scheme provides the most computationally efficient way of calculating \ce{H2O}, \ce{CO}, \ce{CH4}, \ce{N2}, and \ce{NH3} mole fractions, but assumes (a) \ce{H2}-dominated atmosphere with roughly solar elemental abundances, (b) all \ce{C} exists in \ce{CO} and \ce{CH4}, all \ce{O} is in \ce{CO} and \ce{H2O}, and all \ce{N} is in \ce{N2} and \ce{NH3}, (c) the interconversion between these species is defined by two net reactions \ce{CO + 3H2 <=> CH4 + H2O} and \ce{N2 + 3H2 <=> 2NH3}, which make this scheme (d) not applicable under \ce{H2} dissociation ($>$\SI{2500}{\kelvin}) or low pressures. The chemical \textit{equilibrium} scheme computes a local chemical equilibrium using the Gibbs energy minimisation \citep[see][]{Drummond2018}. This scheme calculates mole fractions of a larger number of chemical species under a wider range of elemental abundances and thermodynamical conditions, but assumes chemical equilibrium. Our chemical \textit{relaxation} scheme is similar to that of \citet{Cooper2006}. Our scheme relaxes \ce{CO} and \ce{N2} mole fractions to their equilibrium values at a timescale of the rate-limiting step of \ce{CO}-\ce{CH4} conversion (\ce{H + H2CO + M -> CH3O + M}) and the net \ce{N2}-\ce{NH3} conversion reaction (\ce{N2 + 3H2 <=> 2NH3}), respectively, and computes \ce{CH4}, \ce{H2O} and \ce{NH3} mole fractions from the mass balance assuming all \ce{C} is contained in \ce{CO} and \ce{CH4}, all \ce{O} is in \ce{CO} and \ce{H2O}, and all \ce{N} is in \ce{N2} and \ce{NH3} \citep[see][]{Drummond2018a,Drummond2018b}. The chemical \textit{kinetics} scheme is the most computationally demanding, as it solves stiff ordinary differential equations describing chemical production and loss of chemical species in a chemical network of choice \citep[see][]{Drummond2020} using the DLSODES solver from the Fortran library ODEPACK \citep{Hindmarsh1983}. For the chemical kinetics scheme we use the \citet{Venot2019} chemical network with 30 neutral chemical species and 181 reversible reactions that had been reduced from the larger experimentally validated network of \citet{Venot2012} with the ANSYS\textsuperscript{\tiny\textcopyright} Chemkin-Pro Reaction Workbench\footnote{ANSYS, Inc., San Diego, 2017, Chemkin-Pro 18.2.}, with photodissociation reactions excluded from the reduction. All mentioned chemical networks are available via the KIDA database\footnote{\url{http://kida.obs.u-bordeaux1.fr/}} \citep{Wakelam2012}.

We neglect photochemistry and clouds in this work. Photochemistry generally becomes important at pressures near or less than the lowest included in our modelled domain \citep{Baeyens2022}, while the effects of clouds have been studied in several other works \citep{Lee2016a,Lines2018,Lines2018a,Lines2019,Christie2021,Christie2022,Roman2021}. However, it is important to note that observations do suggest that some hot Jupiter atmospheres are cloud-free \citep{Sing2016,Wakeford2018,Nikolov2018,Alam2021,Sheppard2021,Ahrer2022,Nikolov2022}, so the assumption of a cloud-free atmosphere is a reasonable one for some planets.

\subsection{Model setup and simulation parameters}
\label{section:methods_param}

We adopt the same basic model setup as \citet{Drummond2020}. We use a horizontal resolution of \ang{2.5} longitude by \ang{2} latitude and 66 vertical levels equally spaced in height. Such a choice of grid resolution results in all our simulations being convectively stable, and thus not requiring a dry static adjustment. It is possible that the large-scale flow compensated for transport otherwise performed by unresolved convective or sub-grid scale motions, but this effect would have been minor and localised. We initialise the model at rest with a pressure-temperature profile from the 1D radiative-convective-chemistry model ATMO \citep[e.g.,][]{Drummond2016}. Such a profile, representing a dayside average, is calculated for each planet under the assumption of chemical equilibrium for species present in the \citet{Venot2012} chemical network, and used to derive a column-averaged specific gas constant and specific heat capacity as well as to adjust the vertical extent of the UM to cover a range of pressures from $\sim$\SI{2e7}{\pascal} to $\sim$\SI{1}{\pascal}. Chemical species abundances in simulations using the chemical kinetics scheme are initialised to their chemical equilibrium abundances derived for the pressure-temperature profile described above. Both models (ATMO and the UM) use the same, solar elemental abundances, namely a protosolar value for \ce{He}, meteoritic values for \ce{Li} and \ce{Cs}, and photospheric values for \ce{Na} and \ce{Rb} from \citet{Asplund2009}, and updates for \ce{C}, \ce{N}, \ce{O} and \ce{K} from \citet{Caffau2011}. The UM accounts for the heat flux from the planet interior by adding an upward flux corresponding to an intrinsic temperature of \SI{100}{\kelvin} at the bottom boundary. The choice of having the same (rather than different) intrinsic temperature for all our targets relies on the theory of \citet{Tremblin2017}, which suggests that the hot Jupiter radius inflation problem could be solved by a downward advection of potential temperature rather than a high internal flux, which implies that the choice of intrinsic temperature might be less critical than the simulation length. Since such advection occurs on a timescale of thousands of years \citep{Sainsbury-Martinez2019}, we recognise that our simulations are too short for the deep atmosphere to fully evolve, and hence focus our further analysis on the upper atmosphere only. Additionally, we rely on the finding by \citet[][their Appendix C]{Drummond2020} that HD~209458b simulations with a cold and hot initial pressure-temperature profile showed that the deep atmosphere of this planet does not affect its upper atmosphere dynamics and chemistry, as the quench point for all six chemical species of interest (\ce{H2O}, \ce{CO}, \ce{CO2}, \ce{CH4}, \ce{NH3}, \ce{HCN}) lies above (towards lower pressures) the region of the atmosphere that has not reached a steady-state by the end of the simulations.

The UM uses separate timesteps for the dynamics, radiative transfer, and chemistry, for which we adopt \SI{30}{\second}, \SI{150}{\second} and \SI{3750}{\second}, respectively, as a balance between accuracy and computational efficiency. All simulations are integrated for 1000 Earth days (i.e., \SI{8.64e7}{\second}) providing a pseudo-steady state for the upper atmosphere (diagnosed by analysing the evolution of the global maximum wind velocities, the global mean top-of-the-atmosphere net energy flux and the total mass of opacity sources, see \cref{Appendix_B} for details), while the deeper atmosphere is likely to still be evolving. During the model run time, we use low-resolution spectral files that have 32 spectral bands covering 0.2-322 \si{\micro\metre} and a maximum of 21 k-coefficients per band. For synthetic observations, we use high-resolution spectral files that have 500 spectral bands covering 2-10000 \si{\micro\metre} (to better capture absorption features of chemical species of interest) and a maximum of 15 k-coefficients per band.

Stellar and planetary parameters used to simulate each of our targets with ATMO and the UM are obtained from the TEPCat database\footnote{\url{https://www.astro.keele.ac.uk/jkt/tepcat/}} \citep{Southworth2011} and are summarised in \cref{tab:stellar_params} and \cref{tab:planet_params}, respectively. For the stellar spectrum, we use the Kurucz spectra\footnote{\url{http://kurucz.harvard.edu/stars.html}} for HD~189733 and HD~209458, and the PHOENIX BT-Settl spectra \citep{Rajpurohit2013} for HAT-P-11 and WASP-17.

\begin{table*}
\caption{Stellar parameters used for each simulation.}
\label{tab:stellar_params}
\centering
\setlength\extrarowheight{3pt}
\begin{tabular}{l r r r r r}
\hline\hline
Parameter & Unit & HAT-P-11 & HD~189733 & HD~209458 & WASP-17 \\
\hline
Type                              &                                     & K4             & K1-K2          & G0             & F4 \\
Radius                            & \si{\meter}                         & \num{4.75e+08} & \num{5.23e+08} & \num{8.08e+08} & \num{11.01e+08} \\
Effective temperature             & \si{\kelvin}                        & 4800           & 5050           & 6100           & 6600 \\
Stellar constant at 1 AU          & \si{\watt\per\meter\squared}        & 298.48         & 454.37         & 2335.45        & 5698.12 \\
$\log_{10}$(surface gravity)      & (cgs)                               & 4.50           & 4.53           & 4.38           & 4.00 \\
Metallicity                       & dex                                 & 0.300          & -0.030         & 0.014          & -0.500 \\
\hline
\end{tabular}
\end{table*}

\begin{table*}
\caption{Planetary parameters used for each simulation. Parameters derived from the input stellar and planetary parameters or obtained after the model reached a pseudo-steady state are reported below the dashed line.}
\label{tab:planet_params}
\centering
\setlength\extrarowheight{3pt}
\begin{tabular}{l r r r r r}
\hline\hline
Parameter                             & Unit                                & HAT-P-11b      & HD~189733b     & HD~209458b     & WASP-17b \\
\hline
Inner radius                                & \si{\meter}                         & \num{2.72e+07} & \num{8.05e7}   & \num{9.00e7}   & \num{13.51e7} \\
Domain height              & \si{\meter}                         & \num{0.46e+07} & \num{0.32e7}   & \num{0.90e7}   & \num{3.70e7} \\
Semi major axis                       & AU                                  & 0.05259        & 0.03142        & 0.04747        & 0.05135 \\
Orbital period                        & Earth day                           & 4.888          & 2.219          & 3.525          & 3.735 \\
Rotation rate                         & \si{\radian\per\second}             & \num{1.49e-05} & \num{3.28e-5}  & \num{2.06e-5}  & \num{1.95e-5} \\
Surface gravity\footnotemark[1]       & \si{\meter\per\second\squared}      & 13.20          & 21.50          & 9.30           & 3.31 \\
Specific gas constant                 & \si{\joule\per\kelvin\per\kilogram} & 3514.3         & 3516.1         & 3516.6         & 3518.7 \\
Specific heat capacity                & \si{\joule\per\kelvin\per\kilogram} & \num{1.24e+04} & \num{1.25e4}   & \num{1.28e4}   & \num{1.32e4} \\
\hdashline
Stellar irradiance\footnotemark[2]    & \si{\watt\per\meter\squared}        & \num{0.11e+06} & \num{0.46e+06} & \num{1.04e+06} & \num{2.16e+06} \\
Effective temperature\footnotemark[3] & \si{\kelvin}                        & 813            & 1162           & 1391           & 1622 \\
\hline
\end{tabular}
\\
$^1$Assuming the bottom boundary is at \SI{2e7}{\pascal}.\\
$^2$Calculated as $\mathrm{F_{1 AU}(1/a)^2}$, where $F_{1 AU}$ is the stellar constant at 1 AU, and $a$ is the semi major axis.\\
$^3$Calculated at pseudo-steady state as $\mathrm{(OLR/\sigma)^{1/4}}$, where OLR is the global mean top-of-the-atmosphere \\ outgoing longwave radiation, and $\sigma$ is the Stefan–Boltzmann constant.
\end{table*}

Following \citet{Drummond2020}, we perform a pair of simulations per new target, HAT-P-11b and WASP-17b, with each simulation pair consisting of a simulation using the chemical equilibrium scheme and another using the chemical kinetics scheme. From now on we refer to these simulations as the ``equilibrium'' and ``kinetics'', respectively. These simulations required approximately 5.8 and 8.9 days of wall time per equilibrium and kinetics simulation, respectively, using 216 cores on the DiRAC DIaL supercomputing facility.

\section{Results}
\label{section:results}
Here we describe the dynamical, thermal and chemical structure of the atmospheres of HAT-P-11b, HD~189733b, HD~209458b, and WASP-17b, simulated with the UM using the chemical equilibrium and chemical kinetics scheme. That description is followed by a discussion of the impact of the choice of a chemical scheme on the synthetic observations of these atmospheres.

\subsection{Dynamical structure}
\label{section:dynamical_structure}

\begin{figure*}
\centering
\includegraphics[width=18cm]{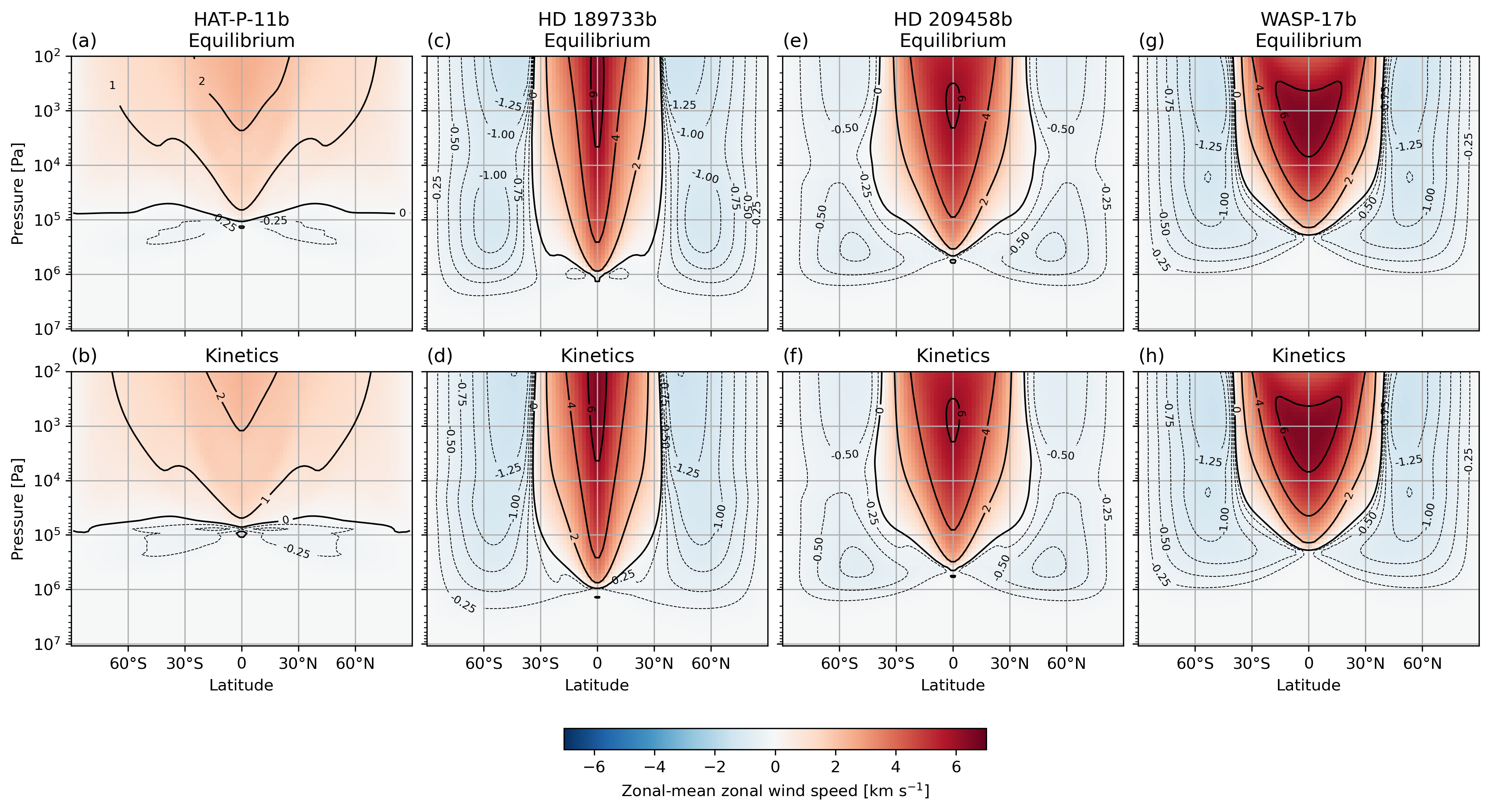}
\caption{Zonal-mean zonal wind speed from HAT-P-11b, HD~189733b, HD~209458b, and WASP-17b equilibrium and kinetics simulations. For the eastward flow, the variability is additionally highlighted in black thick solid contours with \SI{2}{\km\per\second} intervals, and for the westward flow, it is highlighted in black thin dashed contours with \SI{0.25}{\km\per\second} intervals.}
\label{fig:u_znl_mean}
\end{figure*}

Eastward (prograde) zonal flow dominates the atmospheric circulation of HAT-P-11b, HD~189733b, HD~209458b, and WASP-17b (Figures \ref{fig:u_znl_mean}, \ref{fig:temp_hwind_plev_1e4_robinson}-\ref{fig:temp_hwind_plev_1e5_robinson}). The speed and spatial extent of the eastward zonal flow vary from planet to planet, as do the characteristics of the weaker westward (retrograde) zonal flows located deeper than and/or poleward of the eastward flow. While the dynamical structure differs between planets, it is similar between their respective equilibrium and kinetics simulations.

For HAT-P-11b (Figures \ref{fig:u_znl_mean}a, b), the eastward flow extends from pole to pole and down to $\sim$\SI{e5}{\pascal}, and has the maximum zonal-mean zonal wind speed at the equator of \SI{2.6}{\km\per\s} at \SI{1.7e2}{\pascal} in the equilibrium simulation, and \SI{2.4}{\km\per\s} at \SI{1.0e2}{\pascal} in the kinetics simulation. The westward flow prevails at deeper pressures, and a pair of westward zonal jets forms between \numrange{1e5}{3e5} \si{\pascal}, having the zonal-mean zonal wind speed maxima at $\pm$\ang{10} latitude of \SI{0.33}{\km\per\s} in the equilibrium simulation, and \SI{0.62}{\km\per\s} in the kinetics simulation. \citet{Showman2015} predicted a similar atmospheric circulation for a moderately irradiated, slowly rotating hot Jupiter in their ``W$\Omega_{slow}$''\footnote{HD~189733b as $\mathrm{R_{planet}=1.15 \times R_{Jupiter}=\SI{8.04e7}{\meter}}$, $\mathrm{g=\SI{21.4}{\meter\per\square\second}}$, $\mathrm{R_s=\SI{3714.3}{\joule\per\kelvin\per\kilogram}}$, $\mathrm{c_p=\SI{1.30e4}{\joule\per\kelvin\per\kilogram}}$, $\mathrm{metallicity=1 \times solar}$, but with the following parameters modified: $\mathrm{a=0.0789}$ AU, $\mathrm{F_{star}=\SI{7.37e4}{\watt\per\square\second}}$, $\mathrm{T_{orb}=8.8}$ Earth days, $\mathrm{\Omega=\SI{8.264e-6}{\radian\per\second}}$.\label{womega}} cloud-free simulation with the SPARC/MITgcm 3D hydrodynamics-radiation model (their Figures 3 and 4) to that we find for HAT-P-11b. However, their ``W$\Omega_{slow}$'' case is the closest but not equivalent to our HAT-P-11b simulations in terms of the stellar and planetary parameters used.

For HD~189733b (Figures \ref{fig:u_znl_mean}c, d), our model predicts an emergence of the eastward equatorial zonal jet, which is more prominent than that of HAT-P-11b. The HD~189733b equatorial zonal jet spans $\pm$\ang{30} latitude and extends down to $\sim$\SI{e6}{\pascal}, having the maximum zonal-mean zonal wind speed at the equator of \SI{6.4}{\km\per\s} at \SI{3.5e2}{\pascal} in the equilibrium simulation, and \SI{6.5}{\km\per\s} at \SI{2.9e2}{\pascal} in the kinetics simulation. The westward flow controls the circulation elsewhere. Two pair and one pair of westward zonal jets emerge in the equilibrium and kinetics simulations, respectively, but the zonal-mean zonal wind speed maxima between all these westward zonal jets are the same in both simulations, and are equal to \SI{1.5}{\km\per\s}, and are located at $\pm$\ang{40} latitude at $\sim$\SI{e2}{\pascal}.

For HD~209458b (Figures \ref{fig:u_znl_mean}e, f), the emergent eastward equatorial zonal jet is similar to that of HD~189733b, except for being wider, especially between \SIrange{e3}{e5}{\pascal}, and shallower. The HD~209458b equatorial zonal jet spans $\pm$\SI{40}{\degree} latitude and extends down to $\sim$\SI{5e5}{\pascal}, having the maximum zonal-mean zonal wind speed at the equator of \SI{6.1}{\km\per\second} at \SI{8.7e2}{\pascal} in both simulations. The westward flow occurs elsewhere. Two pairs of westward zonal jets emerge in both simulations, with the jet pair located at $\pm$\SI{50}{\degree} latitude at $\sim$\SI{5e5}{\pascal} having faster zonal-mean zonal wind speeds, with maxima of \SI{0.8}{\km\per\second}, than the jet pair located at lower pressures.

For WASP-17b (Figures \ref{fig:u_znl_mean}g, h), our simulations show an eastward equatorial zonal jet similar to those of HD~189733b and HD~209458b, except for it being even shallower. The WASP-17b equatorial zonal jet spans $\pm$\SI{40}{\degree} latitude and extends down to \SI{2e5}{\pascal}, having the maximum zonal-mean zonal wind speed at the equator of \SI{6.6}{\km\per\second} at \SI{1.5e3}{\pascal} in both simulations \citep[which is comparable with \SI{5}{\km\per\second} at $\sim$\SI{e4}{\pascal} estimate from][]{Kataria2016}. The westward flow supports two pairs of zonal jets located at $\pm$\SI{50}{\degree} latitude in both simulations. The jet pair at $\sim$\SI{e3}{\pascal} has faster zonal-mean zonal wind speeds, with maxima of \SI{1.5}{\km\per\second}, than the jet pair located at higher pressures.

\subsection{Thermal structure}
\label{section:thermal_structure}

\begin{figure*}
\centering
  \includegraphics[width=18cm]{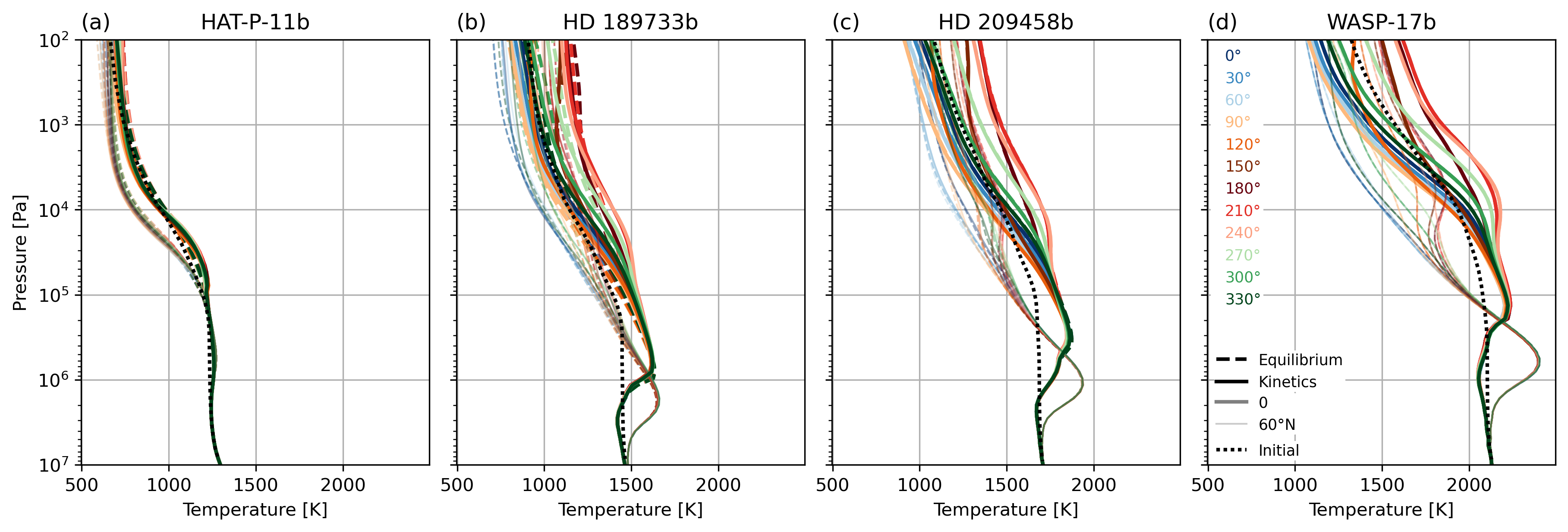}
  \caption{Pressure-temperature profiles for different longitudes around the equator (thick lines in bright colours) and \SI{60}{\degree}N (thin lines in faded colours) from HAT-P-11b, HD~189733b, HD~209458b and WASP-17b equilibrium (dashed) and kinetics (solid) simulations. Pressure-temperature profiles used to initialise each pair of 3D simulations are the dayside average profiles from the respective 1D ATMO equilibrium simulations (dotted).}
  \label{fig:pres_temp_vp_xaxis_same}
\end{figure*}

The thermal structure of the atmospheres of HAT-P-11b, HD~189733b, HD~209458b, and WASP-17b differs between planets but is similar between their respective equilibrium and kinetics simulations. To explore the atmospheric thermal structure of these planets, we analyse their pressure-temperature vertical profiles (Figure \ref{fig:pres_temp_vp_xaxis_same}).

The simulated HAT-P-11b atmosphere is the coldest and the most thermally uniform relative to atmospheres of the other three planets in our sample (Figure \ref{fig:pres_temp_vp_xaxis_same}a). HAT-P-11b receives the lowest stellar irradiance, as it orbits the coldest host star at the largest orbital radius in our sample, resulting in the smallest zonal and meridional temperature gradients in the planet's atmosphere. The HAT-P-11b equilibrium and kinetics simulations predict slightly different thermal structures, especially closer to the equator. At the equator between \num{e2}--\num{e4} \si{\pascal} temperatures are lower by up to \SI{10}{\kelvin} in the kinetics simulation relative to the equilibrium one, while between \num{e4}--\num{e5} \si{\pascal} temperatures are higher by up to \SI{10}{\kelvin} in the kinetics simulation relative to the equilibrium one.

The simulated HD~189733b atmosphere (Figure \ref{fig:pres_temp_vp_xaxis_same}b) is hotter than that of HAT-P-11b. HD~189733b receives a larger stellar irradiance, as it orbits a hotter host star at a smaller orbital radius (in fact, the shortest). This leads to larger zonal and meridional temperature gradients in the planet's atmosphere. The HD~189733b equilibrium and kinetics simulations also predict different thermal structures, but the temperature differences are within 5-10\% \citep{Drummond2020}.

The simulated HD~209458b atmosphere (Figure \ref{fig:pres_temp_vp_xaxis_same}c) is even hotter than that of HD~189733b. HD~209458b receives a larger stellar irradiance, as it orbits a hotter host star but at a larger orbital radius. This results in its zonal and meridional temperature gradients being similar to those of HD~189733b. The HD~209458b equilibrium and kinetics simulations predict similar thermal structures, with the temperature differences being $<$1\% \citep{Drummond2020}.

The simulated WASP-17b atmosphere (Figure \ref{fig:pres_temp_vp_xaxis_same}d) is the hottest, and has the largest zonal and meridional temperature gradients. In fact, its atmosphere is so hot that its equilibrium and kinetics simulations predict almost identical thermal structures due to extremely short chemical timescales.

\subsection{Chemical structure}
\label{section:chemical structure}

\begin{figure*}
\centering
  \includegraphics[width=17cm]{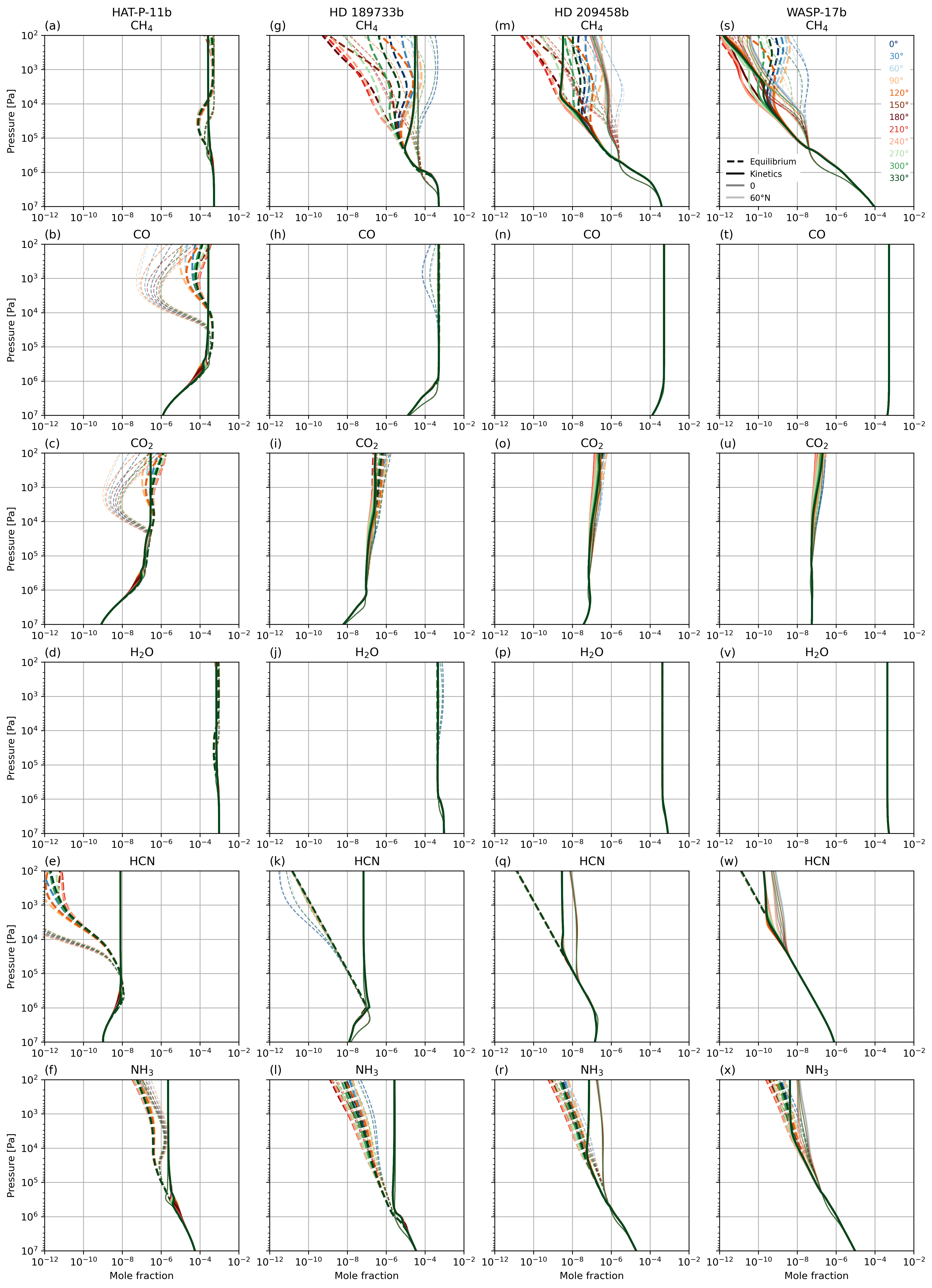}
  \caption{Vertical profiles of chemical species mole fractions for different longitude points around the equator, for the equilibrium (dashed) and kinetics (solid) simulations of HAT-P-11b (first column), HD~189733b (second column), HD~209458b (third column), and WASP-17b (fourth column).}
  \label{fig:pres_chem_vp_xaxis_same}
\end{figure*}

\begin{figure*}
\centering
  \includegraphics[width=18cm]{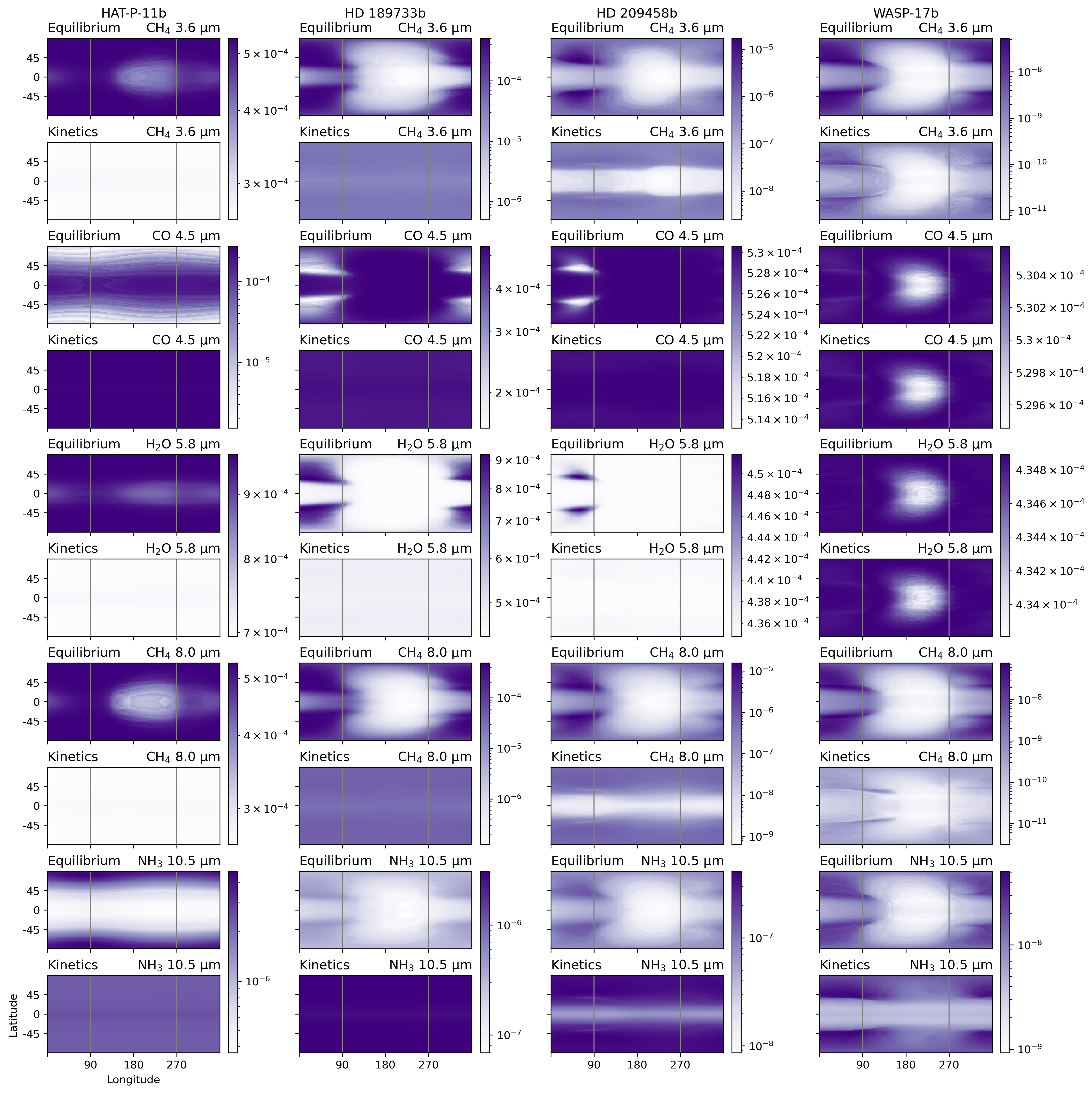}
  \caption{Chemical species mole fractions at a pressure level of the photosphere in individual spectral bands closest to 3.6, 4.5, 5.8, 8.0 and 10.5 \si{\micro\metre} associated with the peaks in absorption from \ce{CH4}, \ce{CO}, \ce{H2O}, \ce{CH4} and \ce{NH3}, respectively, for the equilibrium and kinetics simulations of HAT-P-11b (first column), HD~189733b (second column), HD~209458b (third column), and WASP-17b (fourth column). The substellar point is located at \ang{180} latitude. Grey grid lines show \ang{90} and \ang{270} longitude.}
  \label{fig:chem_at_lw_norm_flux_con_func_of_unity}
\end{figure*}

While the dynamical and thermal structures of the atmospheres of HAT-P-11b, HD~189733b, HD~209458b and WASP-17b are broadly similar between their respective equilibrium and kinetics simulations, their chemical structure is markedly different. To explore the difference in the atmospheric chemical structure between equilibrium and kinetics simulations, we analyse vertical profiles (Figure \ref{fig:pres_chem_vp_xaxis_same}, and Figure \ref{fig:pres_chem_vp_xaxis_diff} with different X-axes) and longitude-latitude cross sections of chemical species mole fractions (Figure \ref{fig:chem_at_lw_norm_flux_con_func_of_unity}). For the longitude-latitude cross sections, we sampled our model data at a pressure level of the photosphere (i.e., where the value of the normalised contribution function is equal to 1, see \cref{Appendix_C}) in spectral bands\footnote{Spectral band centre (in \si{\micro\metre}) and bounds (in \si{\micro\metre}):
\newline \SI{3.6}{\micro\metre} (3.5971223 - 3.6101083),
\newline \SI{4.5}{\micro\metre} (4.484305 - 4.5045044),
\newline \SI{5.8}{\micro\metre} (5.780347 - 5.8139535),
\newline \SI{8.0}{\micro\metre} (8.0 - 8.064516),
\newline \SI{10.5}{\micro\metre} (10.416667 - 10.526316).} closest to the centre of the Spitzer/IRAC channels 3.6, 4.5, 5.8 and 8.0 \si{\micro\metre} that coincide with the peaks in the absorption cross section of \ce{CH4}, \ce{CO}, \ce{H2O} and \ce{CH4}, respectively, and a band closest to 10.5 \si{\micro\metre} coinciding with the peak in \ce{NH3} absorption cross section. We selected a single spectral band for each Spitzer/IRAC channel instead of a range of bands as did \citet{Drummond2018b} to simplify the cases of ``convoluted'' photospheres \citep{Dobbs-Dixon2017}, for which the contribution function has multiple peaks when a range of absorption coefficients from several spectral bands contributes to the total absorption.

The abundances of chemical species in our equilibrium simulations trace the pressure-temperature structure of the atmosphere, as is expected from the Gibbs energy minimisation. In our kinetics simulations, however, they depart from equilibrium due to transport-induced quenching, which is a homogenisation of chemical abundance gradients by atmospheric advection \citep{Moses2014}. Because transport-induced quenching depends on the difference between the timescale of chemical reactions and the timescale of advection, such quenching occurs at different pressure levels, or depths, in the atmosphere for different chemical species \citep{Visscher2011,Agundez2014}. \ce{CH4} and \ce{CO} quenching depends on the timescale of \ce{CH4}-\ce{CO} interconversion and the competition between \ce{CO} and \ce{H2O} for being the main carrier of atomic oxygen \ce{O}, causing \ce{CH4}, \ce{CO} and \ce{H2O} to quench at a similar depth. \ce{CO2} quenching, however, depends on the combined timescale of \ce{CO}-\ce{CO2} conversion and \ce{H2O} destruction, with the latter being the main \ce{OH} source for \ce{CO}+\ce{OH}$\rightarrow$\ce{CO2}+\ce{H}. \ce{CO2} maintains pseudo-equilibrium with \ce{CO} and \ce{H2O} while atmospheric conditions sustain efficient \ce{CO}-\ce{CO2} conversion, but it quenches when it is no longer the case \citep{Tsai2018}. \ce{NH3} and \ce{HCN} quenching is the least understood, but it generally depends on the timescale of \ce{N2}-\ce{NH3} interconversion and \ce{CH4}-\ce{CO} interconversion, with the latter impacting the availability of \ce{C} for a \ce{C}-\ce{N} bond. Because coupled \ce{NH3} and \ce{CH4} chemistry needs to produce \ce{HCN} first \citep{Moses2011}, \ce{NH3} quenches before \ce{HCN} does. Transport-induced quenching also varies in strength, which in our simulations could be thought of as a departure of the disequilibrium abundance not caused by photochemistry (as photochemistry is less relevant at pressures where transport-induced quenching occurs) from equilibrium that depends on the wind speed and direction. When all three types of transport-induced quenching occur in an atmosphere, zonal quenching would likely occur first due to zonal winds being the fastest on average, with meridional and vertical quenching being second and third, respectively. Therefore, capturing the 3D nature of atmospheric flows is important to correctly resolve transport-induced quenching.

\subsubsection{HAT-P-11b}
\label{sec:chem_hatp11b}

For HAT-P-11b, transport-induced quenching occurs at the boundary between the superrotating and the retrograde flow (Figures \ref{fig:u_znl_mean}b, \ref{fig:pres_chem_vp_xaxis_same}a-f). At high latitudes (e.g., \SI{60}{\degree}N), it occurs fully within the retrograde flow, while closer to the equator, it occurs in a calmer but more variable wind regime, which causes the variability in the chemical species mole fractions between \num{e5}--\num{e6} \si{\pascal}.

\ce{CH4}, \ce{CO} and \ce{H2O} quench zonally and vertically, with the depth and strength of either direction of quenching varying with latitude (Figure \ref{fig:pres_chem_vp_xaxis_same}a, b, d). At the equator, \ce{CH4}, \ce{CO} and \ce{H2O} quench at $\sim$\num{e6} \si{\pascal}, where their vertical profiles from the kinetics simulation depart from those at equilibrium. Between \num{e3}-\num{e5} \si{\pascal}, zonal quenching dominates over vertical, as the profiles show almost no change zonally (with longitude) but still change vertically (with decreasing pressure). Between \num{e2}-\num{e3} \si{\pascal}, however, both directions of quenching become equally strong, causing the profiles to stay constant with decreasing pressure until the top of our model domain. At \SI{60}{\degree}N, \ce{CH4}, \ce{CO} and \ce{H2O} quench at the same depth as at the equator, but both quenching directions become equal in strength towards higher pressures, at \num{2e5} \si{\pascal}, than at the equator. Due to these differences in quenching depth and strength with latitude, \ce{CH4}, \ce{CO} and \ce{H2O} mole fractions reach nearly the same value at \num{e3} \si{\pascal}, and form a relatively uniform layer with longitude and latitude between \num{e2}-\num{e3} \si{\pascal}. This contrasts our equilibrium simulation, where they retain zonal and meridional gradients in this layer. The homogenisation of \ce{CH4}, \ce{CO} and \ce{H2O} abundances by atmospheric advection alters their spatial distribution at the pressure level of the photosphere (Figure \ref{fig:chem_at_lw_norm_flux_con_func_of_unity} first column). \ce{CH4} mole fractions at the 3.6 and 8.0 \si{\micro\metre} photospheres decrease by up to a factor of 2 in the kinetics simulation relative to equilibrium one, but stay at $\sim$\num{3e-4} in both simulations. \ce{H2O} mole fractions at the 5.8 \si{\micro\metre} photosphere change similarly to \ce{CH4}, and stay at $\sim$\num{7e-4} in both simulations. \ce{CO} mole fractions at the 4.5 \si{\micro\metre} photosphere, however, increase from $\sim$\num{e-6} to $\sim$\num{e-4} poleward of $\pm$\SI{40}{\degree} latitude, and smooth out the meridional gradient, causing \ce{CO} mole fractions to stay at $\sim$\num{2e-4} at the 4.5 \si{\micro\metre} photosphere in the kinetics simulation.

\ce{CO2} maintains a pseudo-equilibrium with \ce{CO} and \ce{H2O} between \num{1.5e4}--\num{3e5} \si{\pascal}, and quenches zonally and vertically between \num{e2}--\num{1.5e4} \si{\pascal} at a value of $\sim$\num{e-7} (Figure \ref{fig:pres_chem_vp_xaxis_same}c). This differs from our equilibrium simulation, where \ce{CO2} varies with longitude and latitude between \num{e2}--\num{e4} \si{\pascal}, and maintains zonal and meridional gradients of up 2 orders of magnitude.

\ce{NH3} and \ce{HCN} quench in a similar manner to \ce{CH4} and \ce{CO}, but unlike \ce{CH4} and \ce{CO}, their mole fractions mostly increase due to quenching (Figure \ref{fig:pres_chem_vp_xaxis_same}f, e). As a result, \ce{NH3} mole fractions at the 10.5 \si{\micro\metre} photosphere increase in a way that produces a nearly uniform with longitude and latitude \ce{NH3} distribution, with \ce{NH3} mole fractions staying at $\sim$\num{2e-6}, because the meridional gradient between $\sim$\num{4e-6} at the poles and $\sim$\num{4e-7} at the equator present at equilibrium is eliminated.

Results from our HAT-P-11b equilibrium and kinetics simulations for \ce{CH4}, \ce{CO}, \ce{H2O}, \ce{CO2} and \ce{NH3} vertical profiles at the equator agree with those of \citet{Moses2021} from their pseudo-2D chemical equilibrium and kinetics simulations of an exo-Neptune\footnote{$\mathrm{R_{planet}=0.4 \times R_{Jupiter}=\SI{2.80e7}{\meter}}$, $\mathrm{T_{orb}=8.48}$ Earth days,  $\mathrm{\Omega=\SI{8.57e-6}{\radian\per\second}}$, $\mathrm{g=\SI{10}{\meter\per\square\second}}$, $\mathrm{metallicity=1 \times solar}$} with 700 K effective temperature rotating at 0.067294 AU from a K5V star\footnote{$\mathrm{R_{star}=0.7 \times R_{Sun}=\SI{4.87e8}{\meter}}$, $\mathrm{T_{eff}=\SI{4500}{\kelvin}}$, $\mathrm{\log_{10}(g)=4.50 (cgs)}$} (their Figure 6). This agreement supports the prediction made by both their and our study that \ce{H2O}$>$\ce{CH4}$>$\ce{CO} in terms of abundance between \num{e2}-\num{e7} \si{\pascal} in atmospheres of such planets. However, our equatorial \ce{HCN} vertical profiles from the kinetics simulation show mole fractions of $\sim$\num{e-9}, which is 2-3 orders of magnitude lower than those presented in their study. This is likely caused by the difference in \ce{HCN} treatment between \citet{Venot2019} and \citet{Moses2013} chemical networks.

\subsubsection{HD~189733b and HD~209458b}
\label{sec:chem_hd}

Vertical profiles of \ce{CH4}, \ce{CO}, \ce{H2O}, \ce{CO2}, \ce{NH3} and \ce{HCN} were discussed in \citet{Drummond2020}. Here we briefly repeat their major conclusions, and discuss the spatial distribution of \ce{CH4}, \ce{CO}, \ce{H2O} and \ce{NH3} at the pressure level of the photosphere.

For our simulations of HD~189733b and HD~209458b, transport-induced quenching occurs at the base of the superrotating jet, or at higher pressures and within the retrograde flow (Figures \ref{fig:u_znl_mean}d,f, \ref{fig:pres_chem_vp_xaxis_same}g-r).

For HD~189733b, \ce{CH4}, \ce{CO}, \ce{H2O} and \ce{NH3} quench zonally, meridionally and vertically, with the evidence for meridional quenching discussed in \citet{Drummond2018a}. As a result, \ce{CH4} mole fractions at the 3.6 and 8.0 \si{\micro\metre} photospheres are homogenised by atmospheric advection in a way that reduces their zonal and meridional gradients, leaving only a small, $\sim$\num{3e-5} to $\sim$\num{4e-5} equator-to-pole gradient in the kinetics simulation (Figure \ref{fig:chem_at_lw_norm_flux_con_func_of_unity} second column). \ce{CO} and \ce{H2O} mole fractions at the 5.8 \si{\micro\metre} and 4.5 \si{\micro\metre} photospheres, respectively, change mostly on the nightside poleward of $\pm$\SI{30}{\degree} latitude, but with an opposite sign: \ce{H2O} mole fractions decrease by about a factor of 2 and stay at $\sim$\num{4.7e-4}, while \ce{CO} mole fractions increase by about a factor of 2 and stay at $\sim$\num{4.9e-4}. Therefore, \ce{CO}$>$\ce{H2O}$>$\ce{CH4} in terms of abundance in the upper atmosphere of HD~189733b. \ce{NH3} mole fractions at the 10.5 \si{\micro\metre} photosphere change similarly to \ce{CH4}, and stay at $\sim$\num{3e-6} in the kinetics simulation.

For HD~209458b, \ce{CH4}, \ce{CO}, \ce{H2O} and \ce{NH3} quench zonally and vertically. As a result, \ce{CH4} mole fractions at the 3.6 and 8.0 \si{\micro\metre} photospheres are homogenised by atmospheric advection in a way that reduces their zonal gradient, but leaves an equator-to-pole gradient of $\sim$\num{e-9} to $\sim$\num{e-6} in the kinetics simulation (Figure \ref{fig:chem_at_lw_norm_flux_con_func_of_unity} third column), which is larger than that found for HD~189733b. \ce{CO} and \ce{H2O} mole fractions at the 5.8 \si{\micro\metre} and 4.5 \si{\micro\metre} photospheres, respectively, change less than those for HD~189733b and only between 0-\SI{100}{\degree}E poleward of $\pm$\SI{40}{\degree} latitude, but still with an opposite sign: \ce{H2O} mole fractions decrease and stay at $\sim$\num{4.4e-4}, while \ce{CO} mole fractions increase and stay at $\sim$\num{5.3e-4}. Therefore, \ce{CO}$>$\ce{H2O}$>$\ce{CH4} in terms of abundance in the upper atmosphere of HD~209458b. \ce{NH3} mole fractions at the 10.5 \si{\micro\metre} photosphere change similarly to \ce{CH4}, resulting in a small, $\sim$\num{5e-8} to $\sim$\num{4e-7} equator-to-pole gradient in the kinetics simulation.

\subsubsection{WASP-17b}
\label{sec:chem_wasp17b}

For WASP-17b, transport-induced quenching occurs at lower pressures than that for the other three planets. At latitudes occupied by the superrotating jet, it occurs near the fastest part of the jet, while elsewhere it occurs deeper and within the retrograde flow (Figures \ref{fig:u_znl_mean}h, \ref{fig:pres_chem_vp_xaxis_same}s-x).

As WASP-17b's atmosphere is warmer than those of the other three planets, \ce{CO} is thermodynamically favoured over \ce{CH4} in such a way that makes \ce{CO} more abundant than \ce{CH4} throughout our entire model domain. With \ce{H2O} adjusting accordingly, \ce{CO}$>$\ce{H2O}$>$\ce{CH4} in terms of abundance, which is similar to the upper atmospheres of HD~189733b and HD~209458b. Preferential production of \ce{CO} and \ce{H2O} over \ce{CH4} causes \ce{CO} and \ce{H2O} spatial variability to be smaller than that of \ce{CH4}, which is why below we discuss \ce{CH4} vertical profiles only.

\ce{CH4} quenches zonally and vertically, with zonal quenching dominating over vertical, and quenching depth varying with longitude and latitude (Figure \ref{fig:pres_chem_vp_xaxis_same}s). On the dayside at the equator, \ce{CH4} quenches at \num{5e2} \si{\pascal}, while on the nightside at the equator it quenches deeper, with the quenching pressure level increasing from \num{1e3} \si{\pascal} at \SI{270}{\degree}E to \num{7e3} \si{\pascal} at \SI{90}{\degree}E. Therefore, at the equator, \ce{CH4} quenching occurs at different pressure levels for all longitudes. However, at high latitudes, it occurs at a single and higher pressure than that at the equator, which is the same for all longitudes (e.g., at \num{1e5} \si{\pascal} at \SI{60}{\degree}N). This leads to an overall tendency of \ce{CH4} vertical profiles to evolve towards the dayside equilibrium values, which causes \ce{CH4} mole fractions at the 3.6 and 8.0 \si{\micro\metre} photospheres to decrease in a way that retains their meridional gradient but reduces the zonal gradient (Figure \ref{fig:chem_at_lw_norm_flux_con_func_of_unity} fourth column). In the kinetics simulation, \ce{CH4} mole fractions stay at $\sim$\num{e-11} on the dayside and at $\sim$\num{e-10}-\num{e-9} on the nightside. \ce{CO} mole fractions at the 4.5 \si{\micro\metre} photosphere stay at $\sim$\num{5e-4}, while \ce{H2O} mole fractions at the 5.8 \si{\micro\metre} photosphere stay at $\sim$\num{4e-4} in both simulations.

\ce{CO2} is at equilibrium throughout most of our model domain in the kinetics simulation, except for $<$\num{5e2} \si{\pascal} between \SI{330}{\degree}E and \SI{90}{\degree}E on the nightside, where it is likely at pseudo-equilibrium with \ce{CO} and \ce{H2O} (Figure \ref{fig:pres_chem_vp_xaxis_same}u). As a result, \ce{CO2} mole fractions vary with longitude and latitude from $\sim$\num{6e-8} at the substellar point to $\sim$\num{e-7} elsewhere, and the difference between equilibrium and kinetics simulations is small.

\ce{NH3} and \ce{HCN} quench zonally and vertically, with the depth and strength of either direction of quenching varying with longitude and latitude (Figure \ref{fig:pres_chem_vp_xaxis_same}x, w). At the equator, \ce{NH3} quenches at \SI{3e4}{\pascal}, with zonal quenching dominating over vertical between \num{3e3}-\num{3e4} \si{\pascal}, and both directions of quenching becoming equal at lower pressures. At \SI{60}{\degree}N, \ce{NH3} quenches deeper, at \num{7e4} \si{\pascal}, with zonal quenching also dominating over vertical, especially on the nightside, up to the top of our model domain. Zonal quenching is weaker at \SI{60}{\degree}N than at the equator, which causes the \SI{60}{\degree}N profiles for different longitudes to diverge. As a result, \ce{NH3} mole fractions at the 10.5 \si{\micro\metre} photosphere increase, and they increase by a higher amount poleward of $\pm$\SI{40}{\degree} latitude than elsewhere. This creates a more prominent, $\sim$\num{4e-9} to $\sim$\num{e-8} equator-to-pole gradient in the kinetics simulation than at equilibrium. \ce{HCN} spatial variability is similar to that of \ce{NH3}, but \ce{HCN} quenches at lower pressures than \ce{NH3}, at \num{5e3} \si{\pascal} at a mole fraction of $\sim$\num{2e-10} and at \num{3e4} \si{\pascal} at a mole fraction of $\sim$\num{7e-10} at the equator and \SI{60}{\degree}N, respectively.

Results from our WASP-17b equilibrium simulation for \ce{CH4}, \ce{CO}, \ce{H2O} and \ce{NH3} vertical profiles agree with those of \citet{Kataria2016} from their interpolation of 1D pressure-temperature profiles, averaged over the dayside, nightside, east and west terminator, obtained from WASP-17b\footnote{$\mathrm{R_{planet}=1.89 \times R_{Jupiter}=\SI{13.21e7}{\meter}}$, $\mathrm{T_{orb}=3.73}$ Earth days, $\mathrm{\Omega=\SI{1.95e-5}{\radian\per\second}}$, $\mathrm{g=\SI{3.57}{\meter\per\square\second}}$, $\mathrm{metallicity=1 \times solar}$, $\mathrm{F_{star}= \SI{2.14e6}{\watt\per\square\second}}$, $\mathrm{T_{eq}=\SI{1738}{\kelvin}}$} cloud-free simulation of the SPARC/MITgcm 3D hydrodynamics-radiation model onto a chemical equilibrium abundance pressure-temperature grid (their Figures 6, 7; note that instead of mole fractions they show mass mixing ratios, and one needs to multiply the former by about 10 for these species to convert the former to the latter\footnote{The conversion factor from mole fraction to mass fraction is the ratio of the species molar mass to the mean molar mass of the background gas. The latter is about \SI{2.33}{\gram\per\mole} for hot Jupiters. Therefore, the conversion factors are: \ce{CH4} 6.89, \ce{CO} 12.02, \ce{H2O} 7.73, and \ce{NH3} 7.31.}). Such agreement is somewhat surprising, because it happens despite the difference in the jet structure predicted by the UM (one equatorial superrotating jet and two midlatitude counterrotating jets, Figure \ref{fig:u_znl_mean}g) and the SPARC/MITgcm (one pole-to-pole superrotating jet, their Figure 2). This suggests that WASP-17b's thermal structure is mostly controlled by the stellar forcing rather than the atmospheric advection, making this planet a good target for the 1D and 2D modelling studies.

\subsection{Synthetic observations}
\label{section:synthetic_observations}
Here we discuss the impact of the choice between the chemical equilibrium and chemical kinetics scheme on the synthetic observations derived from our simulations of the atmospheres of HAT-P-11b, HD~189733b, HD~209458b, and WASP-17b.

\subsubsection{Transmission spectrum}
\label{section:transmission}

\begin{figure*}
\centering
  \includegraphics[width=18cm]{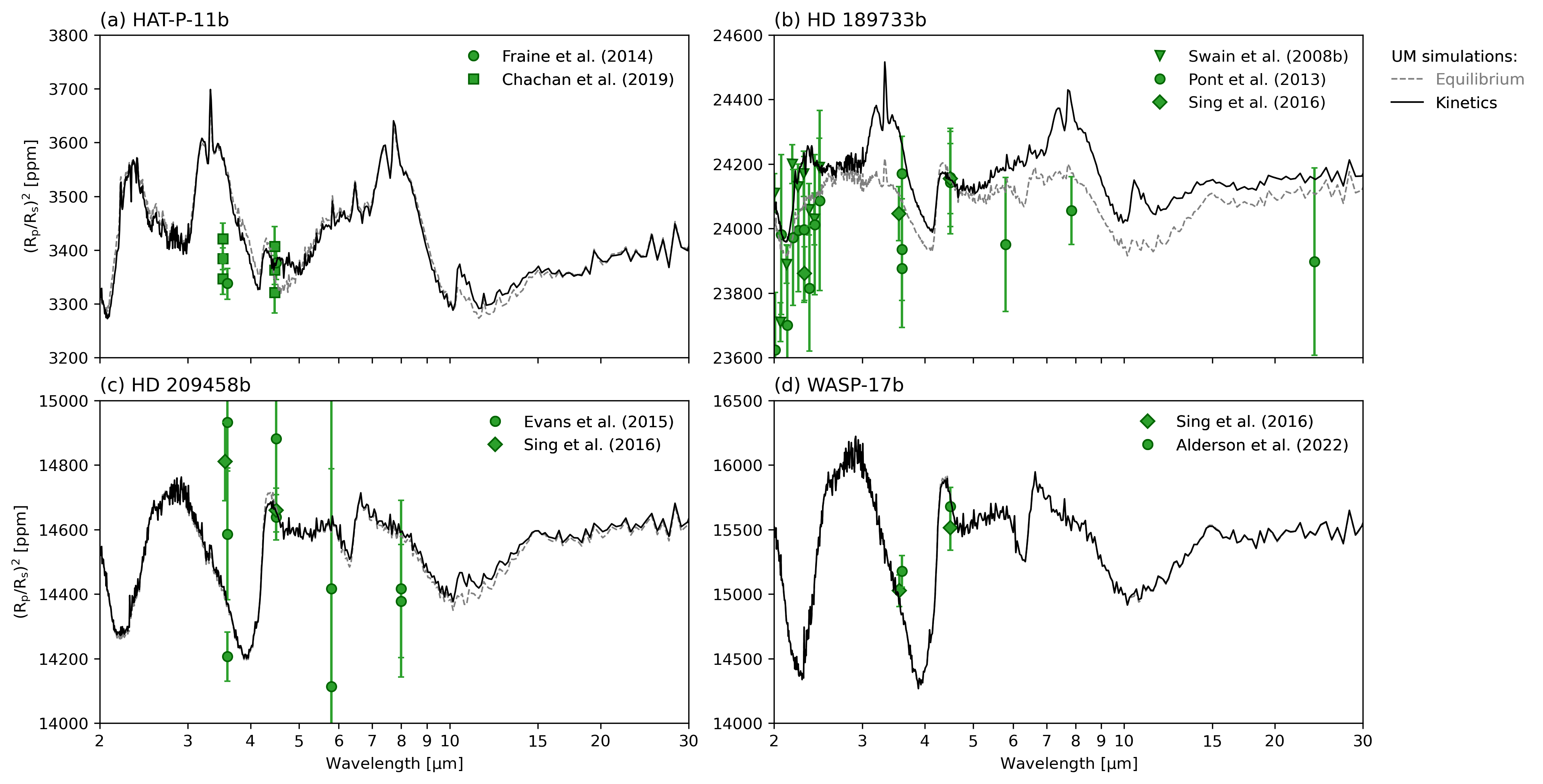}
  \caption{Transmission spectra for the equilibrium (grey dashed) and kinetics (black solid) simulations of HAT-P-11b (a), HD~189733b (b), HD~209458b (c), and WASP-17b (d). UM transmission spectra were uniformly shifted by -760, -980, 140, -7200 ppm, respectively. The observed transit depths and their uncertainties are shown in green.}
  \label{fig:transmission_daynight_total_w_obs}
\end{figure*}

\begin{figure*}
\centering
  \includegraphics[width=18cm]{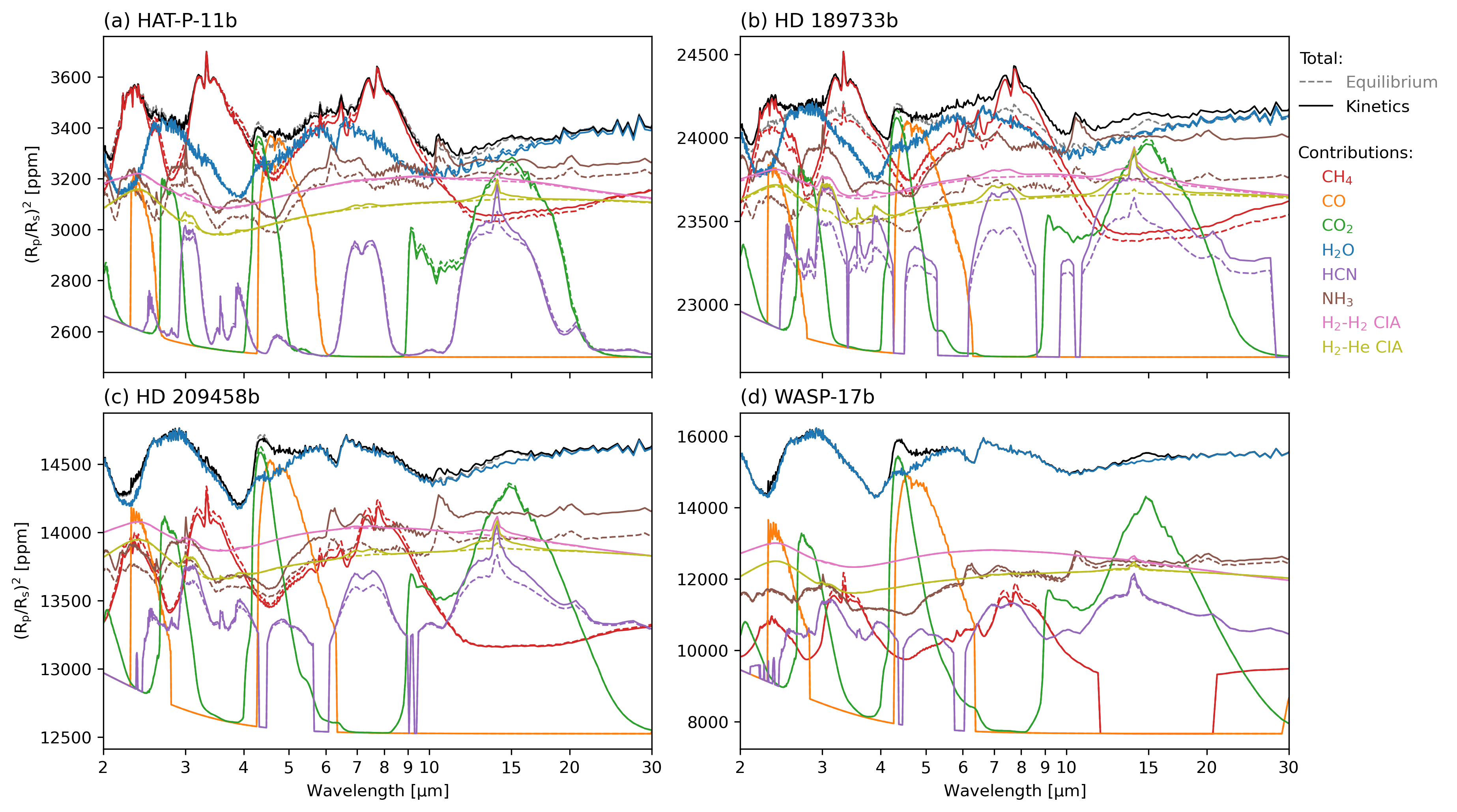}
  \caption{As in Figure \ref{fig:transmission_daynight_total_w_obs} but with the contributions from \ce{CH4}, \ce{CO}, \ce{CO2}, \ce{H2O}, \ce{HCN}, \ce{NH3} absorption, and \ce{H2}-\ce{H2} and \ce{H2}-\ce{He} collision-induced absorption.}
  \label{fig:transmission_daynight_total_w_contributions}
\end{figure*}

Using the methodology of \citet{Lines2018} updated recently by \citet{Christie2021}, we calculate synthetic transmission spectra (Figure \ref{fig:transmission_daynight_total_w_obs}, with differences shown in Figure \ref{fig:transmission_daynight_total_diff}) and the contributions of major opacity sources to these spectra (Figure \ref{fig:transmission_daynight_total_w_contributions}) over the 2-30 \si{\micro\metre} range using the output from our equilibrium and kinetics simulations of the atmospheres of HAT-P-11b, HD~189733b, HD~209458b, and WASP-17b. The observed primary eclipse depth data were taken from the latest publications and observational data reviews, where possible, i.e. for HAT-P-11b, from \citet[Extended Data Table 1]{Fraine2014}, and \citet[Table 2]{Chachan2019}; for HD~189733b, from \citet[Supplementary Table 1]{Swain2008b}, \citet[Table 5]{Pont2013}, and \citet[Figure 1]{Sing2016}; for HD~209458b, from \citet[Table 4]{Evans2015}, and \citet[Figure 1]{Sing2016}; and for WASP-17b, from \citet[Figure 1]{Sing2016}, and \citet[Table 3]{Alderson2022}. Given 1.2-3.3\% uncertainty in the stellar and planetary radii \citep{Southworth2011} used as input into our model, we shifted our synthetic transmission spectra uniformly by a fixed value before comparing them to the observed transit depth data. The value of the shift was chosen so that the latest existing observations at 4.5 \si{\micro\metre} aligned with our model predictions.

Synthetic HAT-P-11b transmission spectra (Figure \ref{fig:transmission_daynight_total_w_obs}a, \ref{fig:transmission_daynight_total_w_contributions}a) from our equilibrium and kinetics simulations differ at several wavelengths. The magnitude and cause of these differences varies, with the largest differences caused by the transport-induced quenching of \ce{CO} and \ce{NH3} (Section~\ref{sec:chem_hatp11b}). In the region probed in transmission in the kinetics simulation, \ce{CO} and \ce{NH3} abundances increase by 2-3 orders of magnitude over those predicted at equilibrium. That causes \ce{CO} and \ce{NH3} to replace \ce{H2O} as the main contributor to the total transmission spectrum at 4.4-5.0 \si{\micro\metre} and 10.2-11.4 \si{\micro\metre}, respectively, and increase the transit depth by up to 60 ppm and 45 ppm, respectively (including smaller contributions from other gases), in the kinetics simulation relative to the equilibrium one. Other, smaller ($\pm$20-30 ppm) differences between our transmission spectra are associated with the transport-induced quenching of \ce{CH4}, \ce{CO2}, \ce{H2O}, \ce{NH3}, and \ce{HCN}. These differences are not dominated by the change in abundance of one gas, but are instead caused by the change in abundance of all of these gases. However, because abundances of these gases often change in opposing directions, the corresponding differences between our spectra are small.

When compared to existing observations, our synthetic HAT-P-11b transmission spectra disagree with observations by \citet{Fraine2014} and \citet{Chachan2019}. However, this disagreement supports the suggestion of \citet{Line2016} and \citet{Chachan2019} that the HAT-P-11b transmission spectrum could be explained by a relatively low atmospheric metallicity and uneven cloud cover. A further comparison with existing observations is required, especially with those shortward of 2 \si{\micro\metre}, but such a comparison is outside the scope of this study.

Synthetic HD~189733b transmission spectra (Figure \ref{fig:transmission_daynight_total_w_obs}b, \ref{fig:transmission_daynight_total_w_contributions}b) from our equilibrium and kinetics simulations depart from each other throughout the entire wavelengths range considered in this work. The main cause of departure is the transport-induced quenching of \ce{CH4}, \ce{NH3}, and \ce{CO2} \citep[][and Section~\ref{sec:chem_hd} in this work]{Drummond2020}. In the region probed in transmission in the kinetics simulation, \ce{CH4} abundance increases by 2-4 orders of magnitude at the eastern terminator, and stays similar to that predicted at equilibrium at the western terminator. This, together with a planet-wide, 2-3 orders of magnitude increase in \ce{NH3} abundance, increases the transit depth by up to 150 ppm at 2.1-2.5 \si{\micro\metre}, 300 ppm at 3.0-4.1 \si{\micro\metre}, and 230 ppm at 7.0-9.5 \si{\micro\metre} (including smaller contributions from other gases) in the kinetics simulation relative to the equilibrium one. Despite the aforementioned increases in \ce{CH4} and \ce{NH3}, the kinetics simulation predicts a smaller (by 30 ppm) transit depth at 4.2-4.5 \si{\micro\metre} than the equilibrium simulation. This is caused by an order of magnitude decrease in \ce{CO2} abundance at the western terminator in the kinetics simulation relative to the equilibrium one. Lastly, and similarly to HAT-P-11b, \ce{NH3} abundance increases in HD~189733b kinetics simulation so that \ce{NH3} replaces \ce{H2O} as the main contributor to the total transmission spectrum at 10.0-11.6 \si{\micro\metre}, and causes an increase the transit depth by up to 200 ppm (including smaller contributions from \ce{H2O}, \ce{CH4}, and \ce{HCN}) at these wavelengths.

When compared to existing observations, our synthetic HD~189733b transmission spectrum from the kinetics simulation partly agrees with observations of \citet{Swain2008b} (reliability of which had been questioned in \citet{Sing2009}), but predicts larger transit depths than those presented in \citet{Pont2013} and \citet{Sing2016}. In fact, in the case of the latter two studies the spectrum from our kinetics simulation overestimates the observed transit depths more severely than the spectrum from our equilibrium simulation. This suggests that if the atmosphere of HD~189733b is cloud- and haze-free, it is closer to being at chemical equilibrium. However, without an observational constraint on the position of the spectrum continuum at the wavelengths range considered in this work, we can not rule out a possibility that HD~189733b's transmission spectrum could be explained by the presence of clouds and/or hazes (as retrieved by, e.g., \citet{Barstow2020}, or simulated with 3D hydrodynamics-radiation model by, e.g., \citet{Steinrueck2021}). As for future observations, \citet{Drummond2020} showed that according to the synthetic JWST and Ariel transmission observations (1 and 10 orbits, respectively), JWST and Ariel could resolve the aforementioned HD 189733b \ce{CH4}-\ce{NH3} and \ce{CO2} spectral features when provided with the data from our equilibrium and kinetics simulations. However, according to the synthetic JWST transmission spectrum, the \ce{NH3} spectral feature at $\sim$10.5 \si{\micro\metre} could be obscured by the noise \citep[][their Figure 12]{Drummond2020}. Overall, it suggests that these telescopes could detect disequilibrium chemistry signatures due to \ce{CH4}-\ce{NH3} and \ce{CO2} in the HD 189733b transmission spectrum, if its atmosphere is cloud- and haze-free.

Synthetic HD~209458b transmission spectra (Figure \ref{fig:transmission_daynight_total_w_obs}c,  \ref{fig:transmission_daynight_total_w_contributions}c) from our equilibrium and kinetics simulations differ at several wavelengths, with the largest differences occurring at 4.1-4.5 \si{\micro\metre} and 10.2-11.3 \si{\micro\metre}. The primary cause of these differences is the transport-induced quenching of \ce{CH4} and \ce{NH3} \citep[][and Section~\ref{sec:chem_hd} in this work]{Drummond2020}. In the region probed in transmission in the kinetics simulation, \ce{CH4} abundance changes in a way that alters the pseudo-equilibrium between \ce{CO}, \ce{H2O}, and \ce{CO2}, and leads to a decrease in \ce{CO2} abundance at the western terminator, which, in turn, decreases the transit depth by up to 37 ppm at 4.1-4.5 \si{\micro\metre} in the kinetics simulation relative to the equilibrium one. Meanwhile, \ce{NH3} abundance increases in the same region, more so at high latitudes than at the equator, and causes an up to 75 ppm increase in the transit depth at 10.2-11.3 \si{\micro\metre}. Other differences between our spectra from kinetics and equilibrium simulations are mostly positive (up to 30 ppm), and are caused by changes in \ce{CH4}, \ce{NH3}, and \ce{HCN} abundances, however, the contribution of these changes to the total transmission spectra are overwhelmed by the contribution of \ce{H2O}.

When compared to existing observations, our synthetic HD~209458b transmission spectra disagree with most of them, although there is substantial variation between the different observational studies. As for future observations, \citet{Drummond2020} showed that according to the synthetic JWST and Ariel transmission observations (1 and 10 orbits, respectively), JWST and Ariel could resolve the aforementioned HD~209458b's \ce{CO2} spectral feature when provided with the data from our equilibrium and kinetics simulations. However, JWST could not resolve HD~209458b's \ce{NH3} spectral feature at $\sim$10.5 \si{\micro\metre} well enough with one orbit \citep[][their Figure 10]{Drummond2020}. Overall, it means that these telescopes could detect a disequilibrium chemistry signature due to \ce{CO2} in the HD~209458b transmission spectrum, if its atmosphere is cloud- and haze-free.

Synthetic WASP-17b transmission spectra (Figure \ref{fig:transmission_daynight_total_w_obs}d,  \ref{fig:transmission_daynight_total_w_contributions}d) from our equilibrium and kinetics simulations depart from each other the most at 4.3-4.4 \si{\micro\metre} and 10.2-11.3 \si{\micro\metre}. Similarly to HD~209458b, the cause of the first departure is the transport-induced quenching of \ce{CH4}, which by altering the pseudo-equilibrium between \ce{CO}, \ce{H2O}, and \ce{CO2}, leads to a decrease in \ce{CO2} abundance at the western terminator (Section~\ref{sec:chem_wasp17b}), which, in turn, causes a decrease in the transit depth by up to 30 ppm in the kinetics simulation relative to the equilibrium one at these wavelengths. Also similarly to HD~209458b, \ce{NH3} abundance increases in WASP-17b kinetics simulation, and causes up to a 20 ppm increase in the transit depth at 10.2-11.3 \si{\micro\metre}. Detecting these disequilibrium chemistry signatures in the WASP-17b's transmission spectrum might be challenging because of the signatures' small size, and the possibility of them to be muted if WASP-17b's atmosphere is not cloud- and haze-free. Other differences between our WASP-17b equilibrium and kinetics spectra are less than $\pm$10 ppm.

When compared to existing observations, our synthetic WASP-17b transmission spectra broadly agree with observations by \citet{Sing2016} and \citet{Alderson2022}. In fact, our simulations show that \ce{H2O} and \ce{CO2} absorption are the main contributors to the total transmission spectra over the 2-30 \si{\micro\metre} range, which compares favourably with the detection of \ce{H2O} (at $>7\sigma$) and the inference of \ce{CO2} (at $>3\sigma$) reported in the latter study. However, a further comparison with observations is required, e.g., with the upcoming JWST observations of WASP-17b transmission spectrum over the 0.6-14 \si{\micro\metre} range (JWST proposal GTO-1353, PI: Nikole Lewis).

\subsubsection{Emission spectrum}
\label{section:emission}

\begin{figure*}
\centering
  \includegraphics[width=18cm]{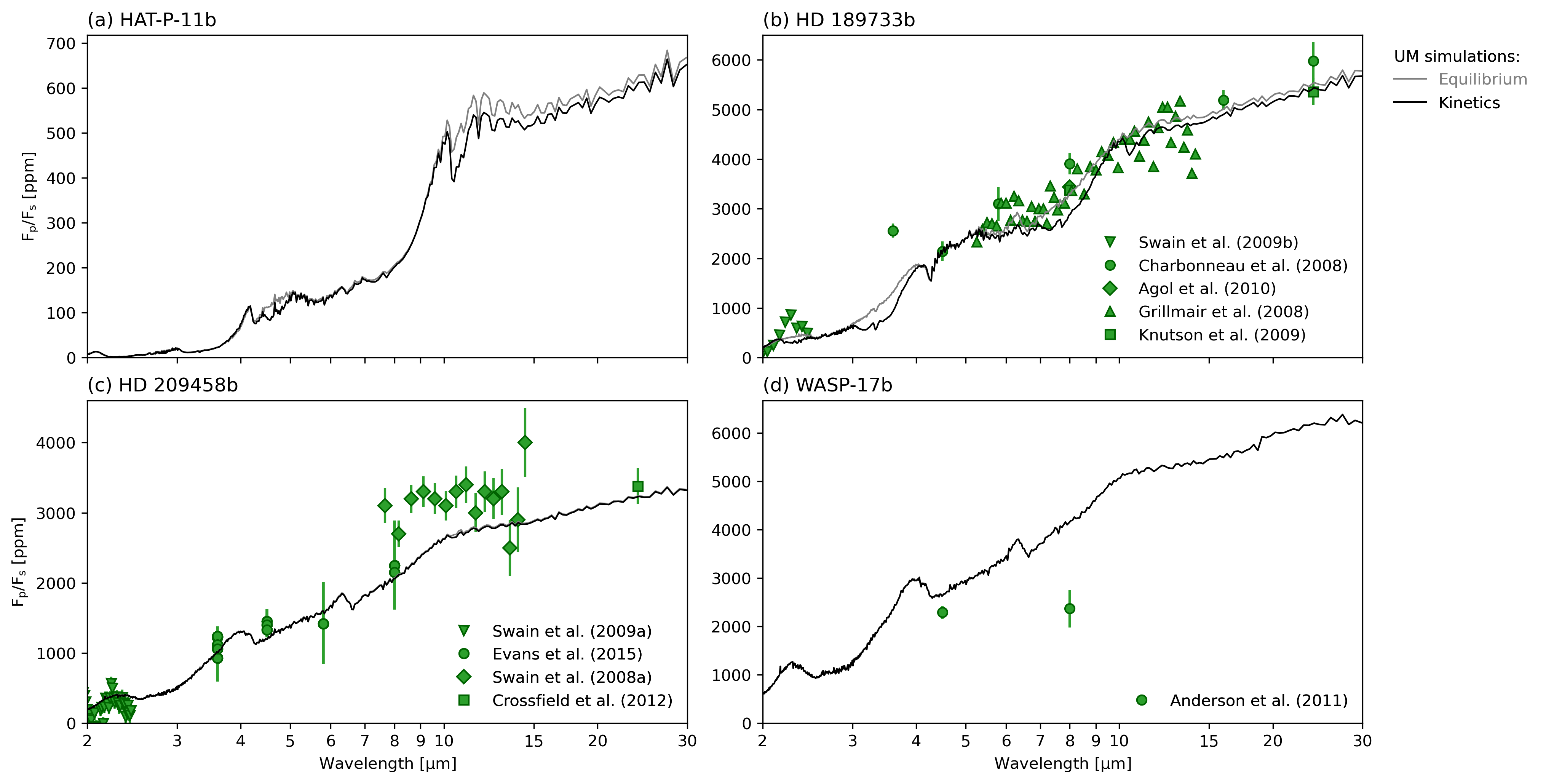}
  \caption{Dayside emission spectra for the equilibrium and kinetics simulations of HAT-P-11b (a), HD~189733b (b), HD~209458b (c), and WASP-17b (d). \citet{Grillmair2008} data were obtained using the WebPlotDigitizer.}
  \label{fig:emission_spectra_w_obs_selection}
\end{figure*}

Following the methodology of \citet{Boutle2017}, we calculate synthetic dayside emission spectra (Figure \ref{fig:emission_spectra_w_obs_selection}) over the 2-30 \si{\micro\metre} range using the output from our equilibrium and kinetics simulations of the atmospheres of HAT-P-11b, HD~189733b, HD~209458b, and WASP-17b. Each synthetic emission spectrum shown here is the average of the spectra at two phase angles, when a maximum area of the planetary dayside is visible to the observer before and after the secondary eclipse. The observed secondary eclipse depth data were taken from the latest publications and observational data reviews, where possible, i.e. for HD~189733b, from \citet[Table 1]{Swain2009a}, \citet[Table 1]{Charbonneau2008}, \citet[Figure 1]{Grillmair2008}, \citet[Section 6.7]{Agol2010}, and \citet[Table 1]{Knutson2009}; for HD~209458b, from \citet[Figure 3]{Swain2009b}, \citet[Table 4]{Evans2015}, \citet[Table 3]{Swain2008a}, and \citet[Table 5]{Crossfield2012}; and for WASP-17b, from \citet[Table 5]{Anderson2011}.

Synthetic HAT-P-11b emission spectra (Figure \ref{fig:emission_spectra_w_obs_selection}a) from our equilibrium and kinetics simulations differ at 4.3-5.0 and 10-30 \si{\micro\metre}. The difference at 4.3-5.0 \si{\micro\metre} is primarily caused by the transport-induced quenching of \ce{CO}, while the difference at 10-30 \si{\micro\metre} is caused by the transport-induced quenching of \ce{NH3} and a temperature decrease (Section~\ref{sec:chem_hatp11b}) with respect to the equilibrium simulation. In the kinetics simulation, a planet-wide increase in \ce{CO} and \ce{NH3} mole fractions at the 4.5 and 10.5 \si{\micro\metre} photospheres, respectively, places these photospheres at lower pressures (Figure \ref{fig:pres_at_lw_norm_flux_con_func_of_unity}), and thus lower temperatures (Figure \ref{fig:temp_at_lw_norm_flux_con_func_of_unity}). This, in turn, causes the emergent intensity at the top of the atmosphere in the 4.5 and 10.5 \si{\micro\metre} spectral bands to decrease, and therefore reduce the planet-to-star flux ratios in the kinetics simulation relative to the equilibrium one. A decrease in temperature, caused by a decrease in the overall absorption due to a decrease in \ce{H2O} mole fractions at pressures lower than \num{e4} \si{\pascal}, additionally reduces the planet-to-star flux ratios at 10-30 \si{\micro\metre} in the kinetics simulation. However, because HAT-P-11b's atmosphere is the coldest relative to the other planetary atmospheres in our sample, the HAT-P-11b emission spectrum has the smallest planet-to-star flux ratios, which make its disequilibrium chemistry signatures due to \ce{CO}, \ce{NH3} and \ce{H2O} challenging to detect.

Synthetic HD~189733b emission spectra (Figure \ref{fig:emission_spectra_w_obs_selection}b) from our equilibrium and kinetics simulations depart from each other at 2.1-2.5, 3.0-4.0, and 5.3-30.0 \si{\micro\metre}. In this case, the main cause of departure is the transport-induced quenching of \ce{CH4} and \ce{NH3} \citep[][and Section~\ref{sec:chem_hd} in this work]{Drummond2020}. An increase in \ce{CH4} mole fractions on the HD~189733b dayside in the kinetics simulation relative to the equilibrium one moves the 3.6 and 8.0 \si{\micro\metre} photospheres on the dayside to lower pressures and temperatures (Figure \ref{fig:pres_at_lw_norm_flux_con_func_of_unity}, \ref{fig:temp_at_lw_norm_flux_con_func_of_unity}), resulting in a decrease in the planet-to-star flux ratios in the  3.6 and 8.0 \si{\micro\metre} spectral bands. Meanwhile, an increase in \ce{NH3} mole fractions on the HD~189733b dayside in the kinetics simulation relative to the equilibrium one does not change the location of the 10.5 \si{\micro\metre} photosphere as much as the change in \ce{CH4} does for the 3.6 and 8.0 \si{\micro\metre} photospheres, causing the emergent intensity at the top of the atmosphere to originate in a layer with relatively similar temperatures to those at chemical equilibrium. Yet, the resulting decrease in the planet-to-star flux ratio in the 10.5 \si{\micro\metre} spectral band in the kinetics simulation is comparable to those for \ce{CH4} due to a high \ce{NH3} absorption cross section at the pressures and temperatures of the 10.5 \si{\micro\metre} photosphere (Figure \ref{fig:absorption_cross_section_relevant_for_hd189733b_kinetics}).

When compared to existing observations, our synthetic HD~189733b emission spectrum from the equilibrium simulation tends to agree more with observations by \citet{Swain2009a}, \citet{Charbonneau2008}, \citet{Grillmair2008}, and \citet{Agol2010}, but less so with observations by \citet{Knutson2009} at 24 \si{\micro\metre} than the kinetics simulation. This suggests that if the atmosphere of HD~189733b is cloud- and haze-free, it is closer to being at chemical equilibrium, or that our kinetics simulation misses some important processes, e.g., formation of clouds, which would generally increase the dayside planetary emission in the infrared part of the spectrum \citep{Christie2021}. As for future observations, \citet{Drummond2020} showed that according to the synthetic JWST and Ariel emission observations (1 and 5 orbits, respectively), JWST and Ariel could resolve the aforementioned HD~189733b \ce{CH4} and \ce{NH3} spectral features when provided with the data from our equilibrium and kinetics simulations \citep[][their Figure 12]{Drummond2020}. That suggests that these telescopes could detect disequilibrium chemistry signatures due to \ce{CH4} and \ce{NH3} in the HD~189733b emission spectrum, if its atmosphere is cloud- and haze-free.

Synthetic HD~209458b emission spectra (Figure \ref{fig:emission_spectra_w_obs_selection}c) from our equilibrium and kinetics simulations differ at 10.2-10.8 \si{\micro\metre}. The main cause of this difference is the transport-induced quenching of \ce{NH3} \citep[][and Section~\ref{sec:chem_hd} in this work]{Drummond2020}. In the kinetics simulation, \ce{NH3} mole fractions increase on the dayside of HD~209458b 10.5 \si{\micro\metre} photosphere, moving this photosphere to slightly lower pressures and temperatures (Figures \ref{fig:pres_at_lw_norm_flux_con_func_of_unity}, \ref{fig:temp_at_lw_norm_flux_con_func_of_unity}), and causing a small decrease in the planet-to-star flux ratios in the 10.5 \si{\micro\metre} spectral band in the kinetics simulation relative to the equilibrium one. HD~209458b's dayside \ce{CH4} distribution at the 3.6 and 8.0 \si{\micro\metre} photospheres does change between equilibrium and kinetics simulations, but this change has a negligible effect on the planet-to-flux ratios in the 3.6 and 8.0 \si{\micro\metre} spectral bands.

When compared to existing observations, our synthetic HD~209458b emission spectra from both equilibrium and kinetics simulations agree reasonably well with observations presented in \citet{Swain2009b}, \citet{Evans2015}, and \citet{Crossfield2012}, but similarly to the results of \citet{Amundsen2016} (their Figure 10) from their cloud- and haze-free HD~209458b UM simulation with the SOCRATES radiation scheme, and the analytical chemical scheme, our simulated planet-to-star flux ratios are smaller relative to those determined by \citet{Swain2008a}. As for future observations, \citet{Drummond2020} showed that according to the synthetic JWST emission observations (1 orbit), JWST could not resolve the aforementioned HD~209458b \ce{NH3} spectral features when provided with the data from our equilibrium and kinetics simulations \citep[][their Figure 10]{Drummond2020}. That suggests that JWST could not detect (at least with 1 orbit) a disequilibrium chemistry signature due to \ce{NH3} in the HD~209458b emission spectrum, if its atmosphere is cloud- and haze-free.

Synthetic WASP-17b emission spectra (Figure \ref{fig:emission_spectra_w_obs_selection}d) from our equilibrium and kinetics simulations are practically the same, suggesting that the emission spectrum of WASP-17b is not affected by transport-induced quenching (Section~\ref{sec:chem_wasp17b}), as the emergent intensity at the top of its atmosphere originates in the part of the atmosphere that is at chemical equilibrium. The disagreement of both our simulations with the existing observations by \citet{Anderson2011} is intriguing, as it suggest that if WASP-17b atmosphere is cloud- and haze-free, our model misrepresents equilibrium chemistry (by assuming, e.g., lower than needed atmospheric metallicity), or misses some processes around the photosphere region. This makes the upcoming JWST observations of WASP-17b emission spectrum over the 0.6-14 \si{\micro\metre} range (JWST proposal GTO-1353, PI: Nikole Lewis) even more welcome.

\subsubsection{Phase curves}
\label{section:phase_curves}

\begin{figure*}
\centering
  \includegraphics[width=18cm]{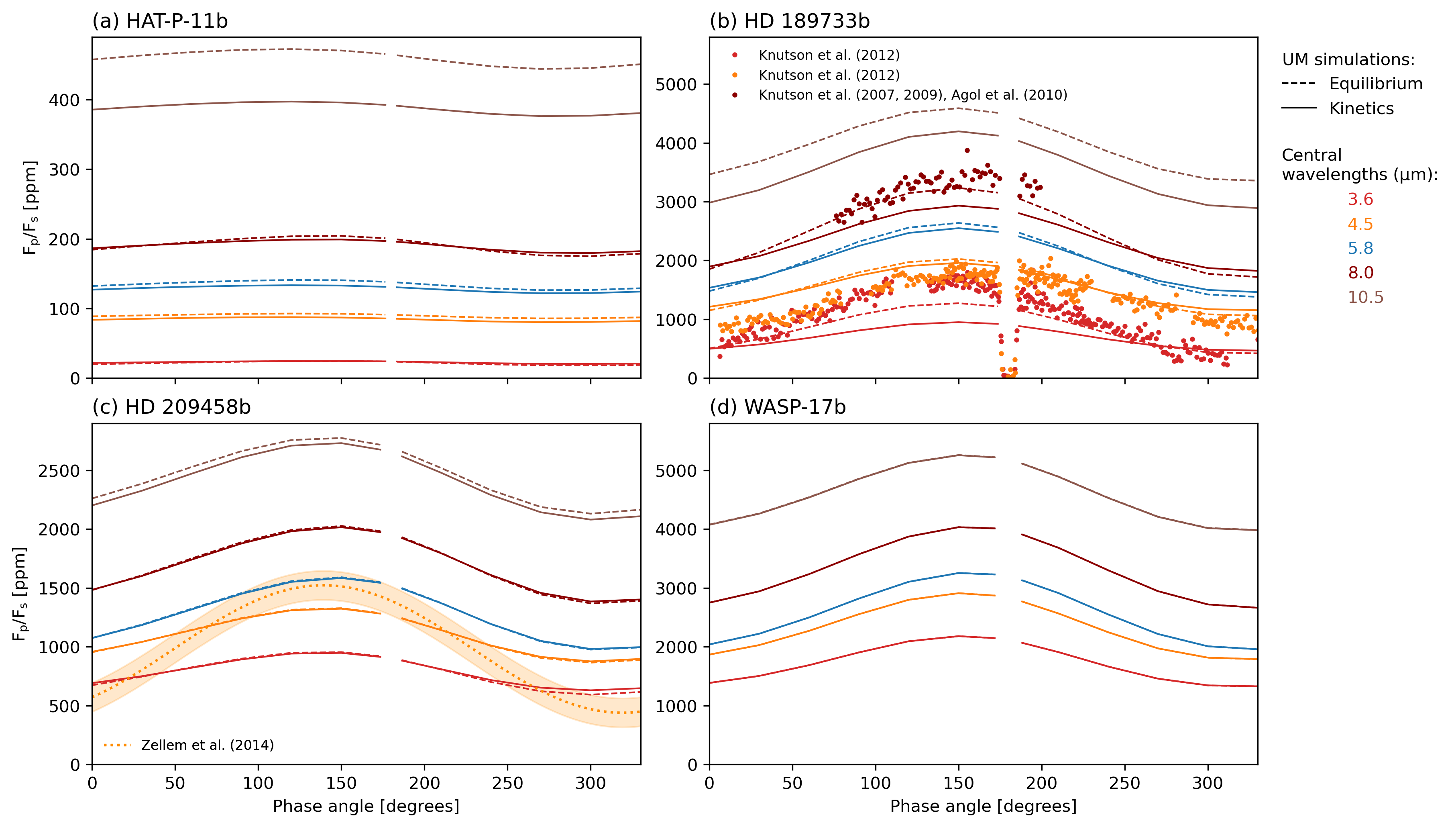}
  \caption{Emission phase curves for the equilibrium (dashed) and kinetics (solid) simulations of HAT-P-11b (a), HD~189733b (b), HD~209458b (c), and WASP-17b (d). The gap in the model data indicates the range of planetary phases when a planet is partially or fully eclipsed by its host star.}
  \label{fig:phase_curves}
\end{figure*}

Following the methodology of \citet{Boutle2017}, we calculate synthetic emission phase curves (Figure \ref{fig:phase_curves}) using the output from our equilibrium and kinetics simulations of the atmospheres of HAT-P-11b, HD~189733b, HD~209458b, and WASP-17b. Each synthetic emission phase curve shown here is comprised of the ratios of planetary to stellar flux summed over the range of SOCRATES spectral bands falling within the corresponding Spitzer/IRAC 3.6, 4.5, 5.8, and 8.0 \si{\micro\metre} bandpasses\footnote{IRAC Instrument Handbook: Spectral Response (\url{https://irsa.ipac.caltech.edu/data/SPITZER/docs/irac/calibrationfiles/spectralresponse/})}, with the exception of 10.5 \si{\micro\metre}, for which we considered only a single SOCRATES spectral band\footnote{Spectral band centre (in \si{\micro\metre}), the corresponding Spitzer/IRAC wavelength bandpass bounds (in \si{\micro\metre}), and the number of SOCRATES spectral bands falling within each considered bandpass:
\newline 3.6 \si{\micro\metre} (3.08106, 4.01038) 76,
\newline 4.5 \si{\micro\metre} (3.72249, 5.22198) 78,
\newline 5.8 \si{\micro\metre} (4.74421, 6.62251) 60,
\newline 8.0 \si{\micro\metre} (6.15115, 9.72875) 60,
\newline 10.5 \si{\micro\metre} (10.416666555101983, 10.52631614584243) 1.}. The observed phase curve data for HD~189733b at 3.6 and 4.5 \si{\micro\metre} were taken from \citet{Knutson2012}, and at 8 \si{\micro\metre} from \citet{Knutson2007,Knutson2009}, and \citet{Agol2010}, while the data for HD~209458b at 4.5 \si{\micro\metre} were taken from \citet{Zellem2014}.

Synthetic HAT-P-11b phase curves (Figure \ref{fig:phase_curves}a) vary little with phase angle, as is expected due to small thermal and chemical gradients predicted for this planet. However, its phase curves still differ between simulations. In the 4.5, 5.8 and 10.5 \si{\micro\metre} bandpasses, phase curves from the kinetics simulation are offset from those from the equilibrium simulation. In the case of 4.5 and 10.5 \si{\micro\metre}, the offset is caused by the transport-induced quenching (Section~\ref{sec:chem_hatp11b}), which by increasing \ce{CO} and \ce{NH3} abundances in a way that evens out their spatial gradients moves the 4.5 and 10.5 \si{\micro\metre} photospheres to lower pressures and temperatures, and decreases planet-to-star flux ratios in these bandpasses. In the case of 5.8 \si{\micro\metre}, however, the offset occurs not because of the transport-induced quenching, but because of a temperature decrease. This decrease in temperature is caused by a decrease in \ce{H2O} abundance at pressures lower then those of the 5.8 \si{\micro\metre} photosphere, which reduces the overall absorption at these pressures, thus reducing the temperature. Lastly, the 3.6 and 8.0 \si{\micro\metre} phase curves from both equilibrium and kinetics simulations are similar, because \ce{CH4} abundance does not change much between simulations.

Synthetic HD~189733b phase curves (Figure \ref{fig:phase_curves}b) vary more with phase angle than those of HAT-P-11b and HD~209458b. While both HD~189733b simulations predict a large day-night thermal gradient, the kinetics simulation predicts smaller chemical gradients, and thus smaller phase curve amplitudes. In the 3.6, 4.5, 5.8 and 8.0 \si{\micro\metre} bandpasses, planet-to-star flux ratios decrease on the dayside but increase on the nightside in the kinetics simulation relative to the equilibrium one. In the case of 3.6, 5.8 and 8.0 \si{\micro\metre}, the differences are caused by the transport-induced quenching (Section~\ref{sec:chem_hd}), which by smoothing \ce{CH4} and \ce{H2O} abundance gradients with longitude flattens phase curves in these bandpasses. In the case of 4.5 \si{\micro\metre}, however, the cause is a combination of changes in abundances of \ce{CO}, \ce{H2O}, and \ce{CO2}, with an increase in \ce{H2O} on the dayside overpowering the contributions from other gases and causing the reduction in flux on the dayside, and a decrease in \ce{H2O} and \ce{CO2} on the nightside controlling the increase in flux on the nightside. Lastly, the 10.5 \si{\micro\metre} phase curve from the kinetics simulation shows consistently lower planet-to-star flux ratios than the equilibrium simulation because of a higher predicted \ce{NH3} abundance.

Comparison of our synthetic HD~189733b phase curves with existing observations produces mixed results. The observed 3.6 and 4.5 \si{\micro\metre} phase curves show similar fluxes before the eclipse (0-180\si{\degree}) but different fluxes after the eclipse (180-360\si{\degree}). Our synthetic 3.6 and 4.5 \si{\micro\metre} phase curves, however, are offset from each other at all phase angles, which agrees with observations after the eclipse but overestimates them at 3.6 \si{\micro\metre} and underestimates them at 4.5 \si{\micro\metre} before the eclipse. If we compare the 3.6 and 8.0 \si{\micro\metre} phase curves, phase curves from our equilibrium simulation tend to agree more with observations than those from our kinetics simulation. This suggests that if HD~189733b atmosphere is cloud- and haze-free, it is closer to being at chemical equilibrium. In fact, if we update our reduced chemical network from that of \citet{Venot2019} to that of \citet{Venot2020}, it will likely increase \ce{CH4} abundance while keeping \ce{CO} abundance the same in our kinetics simulation (see HD~189733b case in \citet{Venot2020} (their Figure 2)). That would, in turn, reduce the fluxes the 3.6 and 8.0 \si{\micro\metre} bandpasses in our kinetics simulation even more, and worsen the agreement with the observed 3.6 and 8.0 \si{\micro\metre} phase curves. Therefore, an alternative direction of model development might be required to improve the agreement between observations and the UM in the case of HD~189733b, e.g., coupling a gas-phase chemistry scheme to a cloud/haze scheme.

Synthetic HD~209458b phase curves (Figure \ref{fig:phase_curves}c) from our equilibrium and kinetics simulations differ in the 3.6 and 8.0 \si{\micro\metre} bandpasses at phase angles corresponding to the nightside, and are offset from each other in the 10.5 \si{\micro\metre} band. The cause of these differences is the transport-induced quenching (Section~\ref{sec:chem_hd}), which by increasing \ce{CH4} abundance on the dayside but decreasing it on the nightside flattens phase curves in the 3.6 and 8.0 \si{\micro\metre} bandpasses, while a planet-wide increase in \ce{NH3} abundance decreases planet-to-star flux ratios in the 10.5 \si{\micro\metre} band in the kinetics simulation relative to the equilibrium one. 

When compared to the fit to phase curve observations in the 4.5 \si{\micro\metre} bandpass \citep{Zellem2014}, both HD~209458b simulations underestimate the observed phase curve amplitude by underestimating the flux on the dayside and overestimating the flux on the nightside. Biases such as these, typical for cloud-free models \citep[e.g.,][]{Parmentier2018}, could be improved upon by including clouds \citep{Parmentier2021}. This was done for HD~209458b with the UM by \citet{Christie2021}, who found that all cloud treatments they considered increase their predicted 4.5 \si{\micro\metre} phase curve amplitude (their Figure 14), but some discrepancies between observations and the model remain.

Synthetic WASP-17b phase curves (Figure \ref{fig:phase_curves}d) vary with phase angle in a similar way as phase curves from our HD~189733b equilibrium simulation. However, in contrast to the other three planets in our sample, WASP-17b's phase curves from both simulations are the same in each considered bandpass, largely because the photospheres corresponding to the centre of each bandpass are located in the region of WASP-17b atmosphere that is at chemical equilibrium (Section~\ref{sec:chem_wasp17b}).

\section{Discussion}
\label{section:discussion}

Transport-induced quenching plays an important role in controlling the chemical composition of the atmospheres of HAT-P-11b, HD~189733b, HD~209458b and WASP-17b. However, how critical this role is in shaping their synthetic spectra and phase curves varies from planet to planet. By expanding the stellar-planetary parameter space explored with our model using both chemical equilibrium and chemical kinetics schemes, we show that transport-induced quenching is predicted to occur in atmospheres of both newly considered planets, a smaller and colder HAT-P-11b and a hotter and larger WASP-17b relative to previously simulated HD~189733b and HD~209458b \citep{Drummond2020}. Yet, when we compare the results for all four planets, the largest differences between synthetic observations produced using our equilibrium and kinetics simulations occur for HD~189733b. It means that there seems to be a ``sweet spot'' combination of stellar and planetary parameters that facilitates the formation of signatures of transport-induced quenching. For the HD~189733 system, such a combination causes our HD~189733b cloud- and haze-free kinetics simulation to have an atmospheric circulation that develops the deepest equatorial jet, and a thermal structure that supports deep quenching, which due to a large contribution from meridional quenching helps shift the photosphere drastically up into the region of lower temperatures. The existence of this ``sweet spot'' emerges as a balance between the timescales of advection and chemistry. In other words, and if we consider the planetary temperature only, to maintain efficient transport-induced quenching, this temperature needs to be low enough to allow a long enough chemical timescale, but high enough so that the thermal gradient between the irradiated hot dayside and the dark cold nightside could sustain a short enough advection timescale.

The observability and detectability of the ``sweet spot'' of transport-induced quenching currently relies heavily on the assumption of a cloud- and haze-free atmosphere. By comparing our model predictions with existing observations, we show that the synthetic observations produced using our kinetics simulations do not necessarily agree better with observations than those produced using our equilibrium simulations. However, the differences in predicted synthetic observations between these simulations (except for WASP-17b emission spectrum and phase curves) are large enough to be identifiable as signatures of transport-induced quenching, and even be detectable with JWST and Ariel (Table \ref{tab:summary}), assuming the atmospheres are cloud- and haze-free.

\begin{table*}
\caption{Observability of signatures of transport-induced quenching. $\square$ signifies the absence or little difference between predictions from our equilibrium and kinetics simulations, $\blacksquare$ indicates the presence of a substantial difference, and the fact that it is potentially detectable with JWST (T1) or Ariel (T2), $\boxplus$ highlights cases where the difference in predicted chemical species abundances is noticeable but unclear if it is detectable, - indicates that no information is available.}
\label{tab:summary}
\centering
\setlength\extrarowheight{3pt}
\begin{tabular}{l l l l l}
\hline\hline
Planet     & Species  & Transit     & Eclipse     & Phase curve \\
\hline
           &          & UM \hspace{0.1cm} T1 T2 & UM \hspace{0.1cm} T1 T2 & UM \\
\hline
HAT-P-11b  & \ce{CH4} & $\square$ \hspace{0.5cm} - - 
                      & $\square$ \hspace{0.5cm} - - 
                      & $\square$ \hspace{0.5cm} - - \\
           & \ce{CO}  & $\blacksquare$ \hspace{0.5cm} - -
                      & $\blacksquare$ \hspace{0.5cm} - - 
                      & $\blacksquare$ \hspace{0.5cm} - - \\
           & \ce{CO2} & $\square$ \hspace{0.5cm} - - 
                      & $\square$ \hspace{0.5cm} - - 
                      & $\square$ \hspace{0.5cm} - - \\
           & \ce{H2O} & $\square$ \hspace{0.5cm} - - 
                      & $\blacksquare$ \hspace{0.5cm} - - 
                      & $\blacksquare$ \hspace{0.5cm} - - \\
           & \ce{HCN} & $\square$ \hspace{0.5cm} - - 
                      & $\square$ \hspace{0.5cm} - - 
                      & $\square$ \hspace{0.5cm} - - \\
           & \ce{NH3} & $\blacksquare$ \hspace{0.5cm} - - 
                      & $\blacksquare$ \hspace{0.5cm} - - 
                      & $\blacksquare$ \hspace{0.5cm} - - \\
\hline
HD~189733b & \ce{CH4} & $\blacksquare$ \hspace{0.5cm} $\blacksquare$ $\blacksquare$
                      & $\blacksquare$ \hspace{0.5cm} $\blacksquare$ $\blacksquare$
                      & $\blacksquare$ \hspace{0.5cm} - - \\
           & \ce{CO}  & $\square$ \hspace{0.5cm} $\square$ $\square$
                      & $\square$ \hspace{0.5cm} $\square$ $\square$
                      & $\blacksquare$ \hspace{0.5cm} - - \\
           & \ce{CO2} & $\blacksquare$ \hspace{0.5cm} $\square$ $\square$
                      & $\square$ \hspace{0.5cm} $\square$ $\square$
                      & $\blacksquare$ \hspace{0.5cm} - - \\
           & \ce{H2O} & $\square$ \hspace{0.5cm} $\square$ $\square$
                      & $\square$ \hspace{0.5cm} $\square$ $\square$
                      & $\blacksquare$ \hspace{0.5cm} - - \\
           & \ce{HCN} & $\square$ \hspace{0.5cm} $\square$ $\square$
                      & $\square$ \hspace{0.5cm} $\square$ $\square$
                      & $\square$ \hspace{0.5cm} - - \\
           & \ce{NH3} & $\blacksquare$ \hspace{0.5cm} $\boxplus$ - 
                      & $\blacksquare$ \hspace{0.5cm} $\blacksquare$ -
                      & $\blacksquare$ \hspace{0.5cm} - - \\
\hline
HD~209458b & \ce{CH4} & $\square$ \hspace{0.5cm} $\square$ $\square$ 
                      & $\square$ \hspace{0.5cm}  $\square$ $\square$
                      & $\square$ \hspace{0.5cm} - - \\
           & \ce{CO}  & $\square$ \hspace{0.5cm} $\square$ $\square$ 
                      & $\square$ \hspace{0.5cm} $\square$ $\square$
                      & $\square$ \hspace{0.5cm} - - \\
           & \ce{CO2} & $\blacksquare$ \hspace{0.5cm} $\blacksquare$ $\blacksquare$
                      & $\square$ \hspace{0.5cm} $\square$ $\square$
                      & $\square$ \hspace{0.5cm} - - \\
           & \ce{H2O} & $\square$ \hspace{0.5cm} $\square$ $\square$
                      & $\square$ \hspace{0.5cm} $\square$ $\square$
                      & $\square$ \hspace{0.5cm} - - \\
           & \ce{HCN} & $\square$ \hspace{0.5cm} $\square$ $\square$
                      & $\square$ \hspace{0.5cm} $\square$ $\square$
                      & $\square$ \hspace{0.5cm} - - \\
           & \ce{NH3} & $\blacksquare$ \hspace{0.5cm} $\boxplus$ - 
                      & $\blacksquare$ \hspace{0.5cm} $\blacksquare$ -
                      & $\blacksquare$ \hspace{0.5cm} - - \\
\hline
WASP-17b   & \ce{CH4} & $\square$ \hspace{0.5cm} - -
                      & $\square$ \hspace{0.5cm} - -
                      & $\square$ \hspace{0.5cm} - -\\
           & \ce{CO}  & $\square$ \hspace{0.5cm} - -
                      & $\square$ \hspace{0.5cm} - -
                      & $\square$ \hspace{0.5cm} - -\\
           & \ce{CO2} & $\blacksquare$ \hspace{0.5cm} - -
                      & $\square$ \hspace{0.5cm} - -
                      & $\square$ \hspace{0.5cm} - -\\
           & \ce{H2O} & $\square$ \hspace{0.5cm} - -
                      & $\square$ \hspace{0.5cm} - -
                      & $\square$ \hspace{0.5cm} - -\\
           & \ce{HCN} & $\square$ \hspace{0.5cm} - - 
                      & $\square$ \hspace{0.5cm} - -
                      & $\square$ \hspace{0.5cm} - -\\
           & \ce{NH3} & $\blacksquare$ \hspace{0.5cm} - -
                      & $\square$ \hspace{0.5cm} - -
                      & $\square$ \hspace{0.5cm} - -\\
\hline
\end{tabular}
\end{table*}

Lastly, and despite that our sample of planets is small, we would like to report evidence for the trend in chemical abundances that depends on the effective planetary temperature. In our equilibrium and kinetics simulations, \ce{CH4}, \ce{H2O}, \ce{NH3} and \ce{HCN} decrease, \ce{CO2} stays relatively constant and \ce{CO} increases with increasing effective planetary temperature. This agrees with the results of \citet{Baeyens2021} from a relevant subset of pseudo-2D chemical kinetics simulations of planets with 800-1600 K effective planetary temperature rotating around K5, G5 and F5 stars (their Figures 8 and E9).

\section{Conclusions}
\label{section:conclusions}

We present results from a set of cloud- and haze-free simulations of the atmospheres of HAT-P-11b, HD~189733b, HD~209458b, and WASP-17b computed using a coupled 3D hydrodynamics-radiation-chemistry model, the Met Office Unified Model (UM). This work expands the stellar-planetary parameter space explored with the UM relative to the similar previous work of \citet{Drummond2020}.

We conclude that according to our equilibrium and kinetics simulations:
\begin{enumerate}
\item Transport-induced quenching occurs in atmospheres of all planets considered in our work, however the extent to which transport-induced quenching affects their synthetic observations varies from planet to planet. 
\item Due to a combination of stellar and planetary parameters, HD~189733b's atmosphere develops a particular balance between its advection and chemistry timescales, that places it at the “sweet spot” for identification of signatures of transport-induced quenching.
\item However, this “sweet spot” emerges when we assume a cloud- and haze-free atmosphere, which might be a good assumption for some planets, e.g., WASP-96b \citep{Nikolov2018,Nikolov2022}, making such planets the best targets for testing this theory, but when it is not a fair assumption, the observability and detectability of signatures of transport-induced quenching should, if possible, be re-evaluated by taking into account processes that affect them.
\end{enumerate}

\subsection{Future work}
\label{section:future_work}

The potential to detect signatures of disequilibrium chemistry in general is likely to be larger when both transport-induced quenching and another disequilibrium process, photochemistry \citep{Moses2014}, are considered together. Photochemistry was shown to have an impact on the gas-phase composition of \ce{H2}-dominated atmospheres both experimentally \citep{Fleury2019,Fleury2020} and through modelling \citep[most recently by][]{Baeyens2022}. While \citet{Baeyens2022} found that photochemistry has a limited impact on the synthetic transmission spectra of tidally-locked \ce{H2}-dominated planets, they considered photochemistry in isolation from photochemical hazes. Pathways of formation of photochemical hazes in atmospheres of warm and hot Jupiters are not yet known, as laboratory data for the relevant gas mixtures and temperatures have only recently started to be obtained \citep{Fleury2019,He2020}. However, formation of aerosols (hazes particles) from initial gas mixtures without the need for artificial nucleation centres was reported already, which confirms the existence of a transition from gas-phase only homogeneous chemistry to gas-solid heterogeneous chemistry in such regimes. This highlights the need for a good representation of (1) chemistry in the absence of light (i.e., without an interaction with a photon), (2) photochemistry (i.e., with an interaction with a photon), (3) transformation of gas into solid, and (4) gas-solid heterogeneous chemistry. Given the complexity of photochemical haze formation and computational expense of 3D modelling, in the short-term we plan to address (1) by implementing the update to the methanol chemistry recommended for warm Neptune simulations by \citet{Venot2020} and/or use the \citet{Tsai2022} chemical network, and (2) by leveraging the work done by \citet{Ridgway2022}, who added the treatment of photodissociation reactions into the UM, leaving (3) and (4) for the medium- to long-term.

Another medium- to long-term goal is to couple the UM's chemical kinetics scheme to a cloud formation scheme. Previously, the UM's analytic chemical scheme based on \citet{Burrows1999} had been coupled by \citet{Lines2018a} to the kinetic, microphysical mineral cloud formation model of \citet{Lee2016a} and by \citet{Lines2019} and \citet{Christie2021,Christie2022} to the EDDYSED cloud model of \citet{Ackerman2001}. Coupling the UM's chemical kinetics scheme to any cloud scheme would require an extension of the chemical network to include cloud forming species relevant for \ce{H2}-dominated atmospheres.

Lastly, according to our simulations, transport-induced quenching occurs deep within the atmospheres of HAT-P-11b and HD~189733b. It suggests that capturing interactions between the planetary interior and the atmosphere above is likely to be particularly important for these two planets, and we would like to simulate that part of the atmosphere better in the future.

\section*{Acknowledgements}
The authors thank the anonymous referee for a thorough review of our paper. We also thank Maria E. Steinrueck and Heather A. Knutson for providing the observed HD~209458b phase curve data published in \citet{Knutson2012}, and Denis Sergeev for help with the data visualisation. This research has made use of the NASA Exoplanet Archive, which is operated by the California Institute of Technology, under contract with the National Aeronautics and Space Administration under the Exoplanet Exploration Program.\\
Financial support. This research was partly supported by a Science and Technology Facilities Council Consolidated Grant ST/R000395/1a, a UKRI Future Leaders Fellowship MR/T040866/1, and the Leverhulme Trust through a research project grant RPG-2020-82. JM acknowledges the support of a Met Office Academic Partnership secondment.\\
Hardware. This research was performed using the DiRAC Data Intensive service at Leicester, operated by the University of Leicester IT Services, which forms part of the STFC DiRAC HPC Facility (\url{www.dirac.ac.uk}). The equipment was funded by BEIS capital funding via STFC capital grants ST/K000373/1 and ST/R002363/1 and STFC DiRAC Operations grant ST/R001014/1. DiRAC is part of the National e-Infrastructure.\\
Software. Material produced using Met Office Software. The scripts to process and visualise the Met Office Unified Model data are available on GitHub at \url{https://github.com/mzamyatina/signatures_of_wind_driven_chemistry_on_hot_jupiters}; these scripts are dependent on the following Python libraries: aeolus \citep{aeolus}, iris \citep{iris}, ipython \citep{ipython}, jupyter \citep{jupyternotebook}, numpy \citep{numpy}, pandas \citep{pandas}, and matplotlib \citep{matplotlib}.\\
We would also like to note the energy intensive nature of supercomputing, especially simulations with interactive chemistry. We estimate the final production runs needed for this paper resulted in roughly 4.5 t\ce{CO2}e emitted into the atmosphere.

\section*{Data availability}
The research data supporting this publication are openly available from the Open Research Exeter (ORE) online repository at: \url{https://doi.org/10.24378/exe.4304}. For the purpose of open access, the authors have applied a Creative Commons Attribution (CC BY) licence to any Author Accepted Manuscript version arising.

\bibliographystyle{mnras}
\bibliography{references}

\appendix

\section{Line lists}
\label{Appendix_A}

\begin{table*}
\caption{Line lists and partition functions used in this work. More details can be found in \citet{Goyal2020}.}
\label{tab:line_lists}
\centering
\setlength\extrarowheight{3pt}
\begin{tabular}{l l l }
\hline\hline
Species & Line list & Partition function \\
\hline
\ce{Li}             & VALD3\footnotemark[1]         & \citet{Sauval1984} \\
\ce{Na}             & VALD3\footnotemark[1]         & \citet{Sauval1984} \\
\ce{K}              & VALD3\footnotemark[1]         & \citet{Sauval1984} \\
\ce{Rb}             & VALD3\footnotemark[1]         & \citet{Sauval1984} \\
\ce{Cs}             & VALD3\footnotemark[1]         & \citet{Sauval1984} \\
\ce{H2O}            & \citet{Barber2006}            & \citet{Barber2006} \\
\ce{CO}             & \citet{Rothman2010}           & \citet{Rothman2009} \\
\ce{CO2}            & \citet{Tashkun2011}           & \citet{Rothman2009} \\
\ce{CH4}            & \citet{Yurchenko2014}         & \citet{Yurchenko2014} \\
\ce{NH3}            & \citet{Yurchenko2011}         & \citet{Sauval1984} \\
\ce{HCN}            & \citet{Harris2006,Barber2014} & \citet{Harris2006,Barber2014} \\
\ce{H2}-\ce{H2} CIA & \citet{Richard2012}           & N/A \\
\ce{H2}-\ce{He} CIA & \citet{Richard2012}           & N/A \\
\hline
\end{tabular}
\\
$^1$\citet{Heiter2008,Heiter2015,Ryabchikova2015} (\url{http://vald.astro.uu.se/~vald/php/vald.php})
\end{table*}

\begin{table*}
\caption{Pressure broadening data used in this work. More details can be found in \citet{Goyal2020}.}
\label{tab:broadeners}
\centering
\setlength\extrarowheight{3pt}
\begin{tabular}{l l l l}
\hline\hline
Species & Broadener & Line width & Exponent \\
\hline
\ce{Li}, \ce{Rb}, \ce{Cs} & \ce{H2} & \citet{Allard1999}                       & \citet{Sharp2007}\\
                          & \ce{He} & \citet{Allard1999}                       & \citet{Sharp2007}\\
\ce{Na}, \ce{K}           & \ce{H2} & \citet{Allard1999,Allard2003,Allard2007} & \citet{Sharp2007}\\
                          & \ce{He} & \citet{Allard1999,Allard2003,Allard2007} & \citet{Sharp2007}\\
\ce{H2O}                  & \ce{H2} & \citet{Gamache1996}                      & \citet{Gamache1996} \\
                          & \ce{He} & \citet{Solodov2009,Steyert2004}          & \citet{Gamache1996} \\
\ce{CO}                   & \ce{H2} & \citet{Regalia-Jarlot2005}               & \citet{LeMoal1986} \\
                          & \ce{He} & \citet{BelBruno1982,Mantz2005}           & \citet{Mantz2005} \\
\ce{CO2}                  & \ce{H2} & \citet{Padmanabhan2014}                  & \citet{Sharp2007} \\
                          & \ce{He} & \citet{Thibault1992}                     & \citet{Thibault2000} \\
\ce{CH4}                  & \ce{H2} & \citet{Pine1992,Margolis1993}            & \citet{Margolis1993} \\
                          & \ce{He} & \citet{Pine1992}                         & \citet{Varanasi1990} \\
\ce{NH3}                  & \ce{H2} & \citet{Hadded2001,Pine1993}              & \citet{Nouri2004} \\
                          & \ce{He} & \citet{Hadded2001,Pine1993}              & \citet{Sharp2007} \\
\ce{HCN}                  & \ce{H2} & \citet{Landrain1997}                     & \citet{Sharp2007} \\
                          & \ce{He} & \citet{Landrain1997}                     & \citet{Sharp2007} \\
\hline
\end{tabular}
\end{table*}

\section{Model pseudo-steady state}
\label{Appendix_B}

\begin{figure*}
\centering
  \includegraphics[width=18cm]{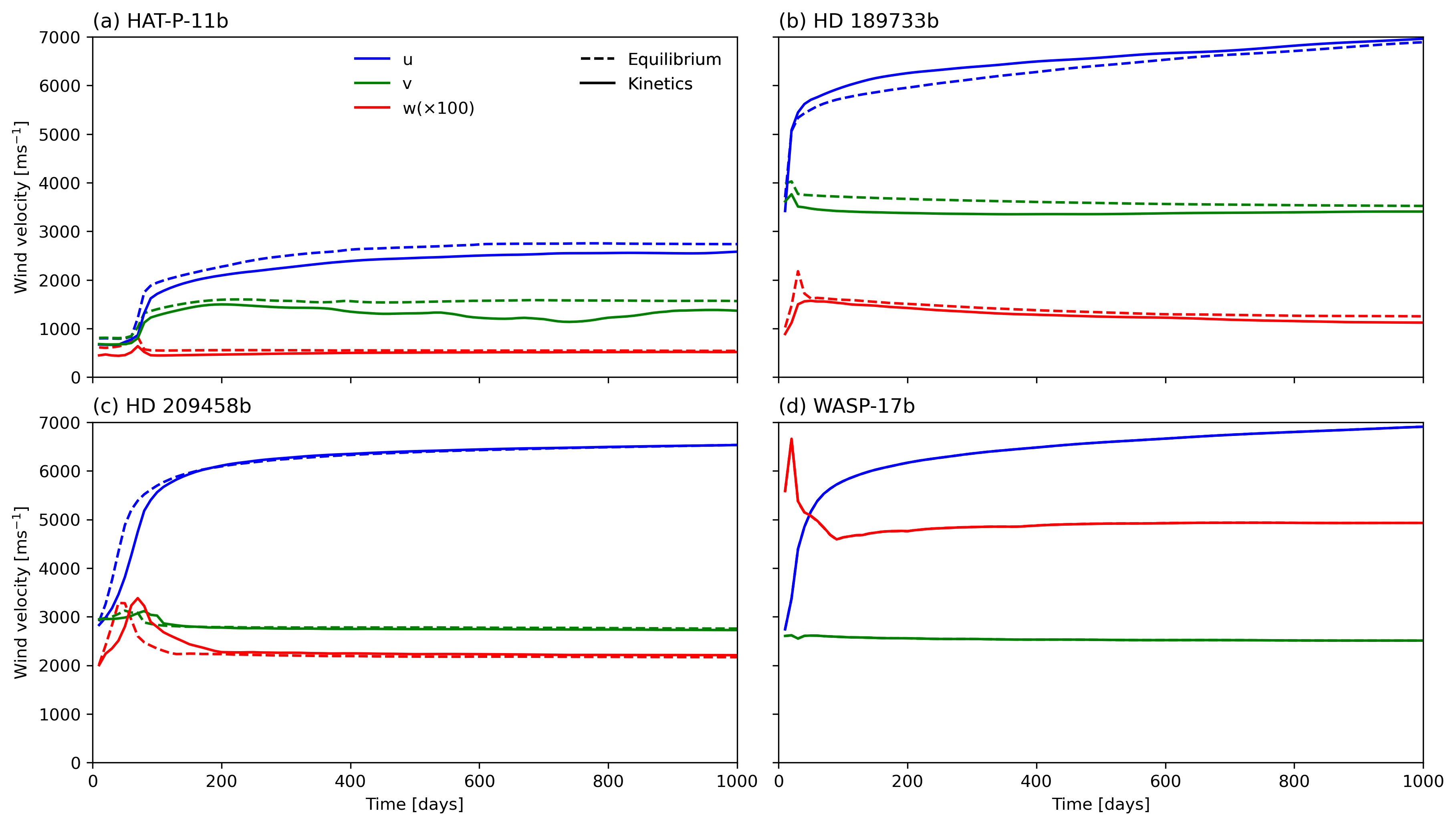}
  \caption{Evolution of the global maximum of the zonal (blue), meridional (green) and vertical (red) components of the wind velocity from the equilibrium (dashed) and kinetics (solid) simulations of HAT-P-11b (a), HD~189733b (b),  HD~209458b (c) and WASP-17b (d). Note that the vertical wind velocity is scaled by a factor of a 100.}
  \label{fig:steady_state_ts_wind_max}
\end{figure*}

\begin{figure*}
\centering
  \includegraphics[width=18cm]{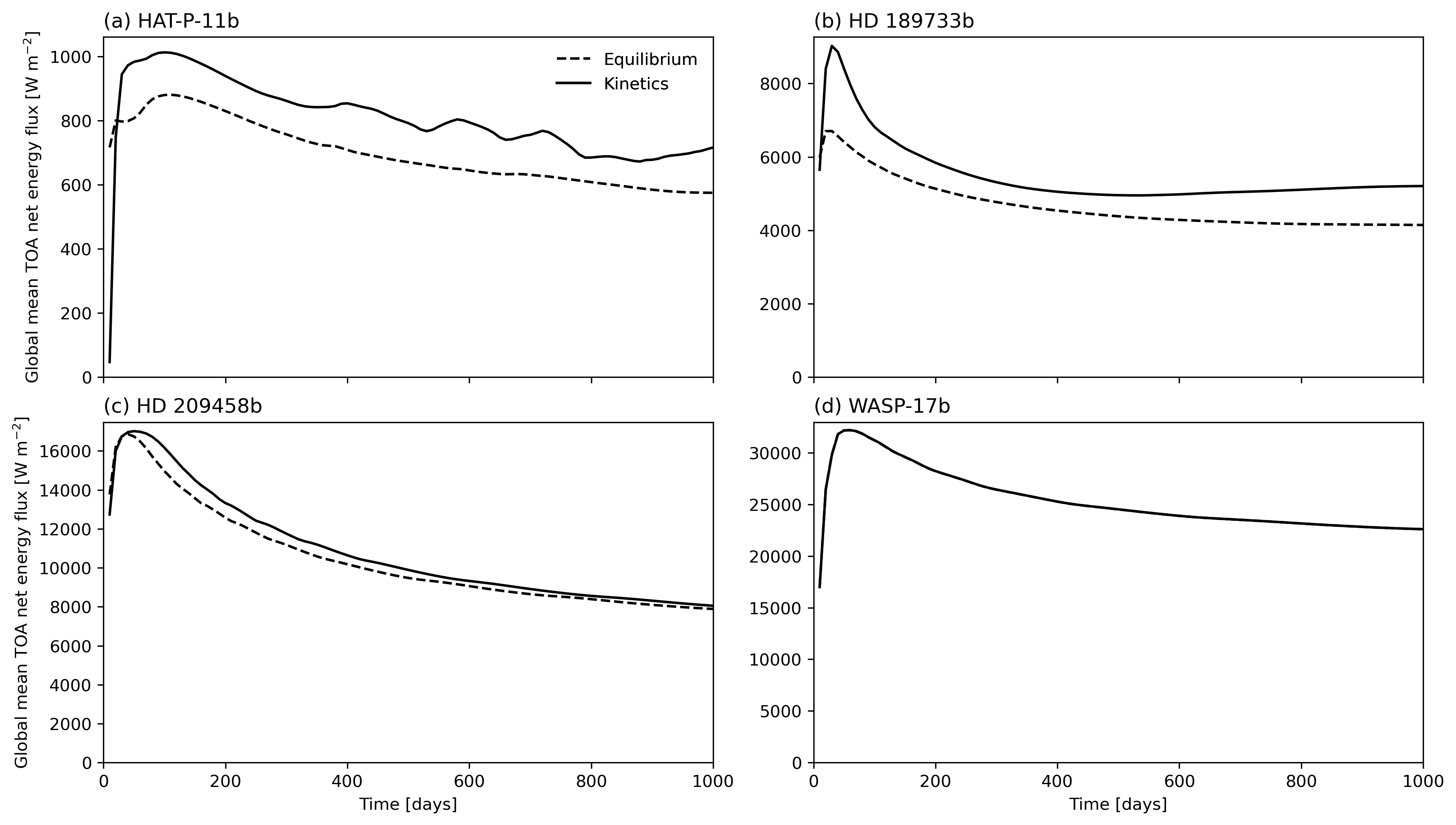}
  \caption{Evolution of the global mean top-of-atmosphere net energy flux from the equilibrium (dashed) and kinetics (solid) simulations of HAT-P-11b (a), HD~189733b (b), HD~209458b (c) and WASP-17b (d).}
  \label{fig:steady_state_ts_toa_net}
\end{figure*}

\begin{figure*}
\centering
  \includegraphics[width=18cm]{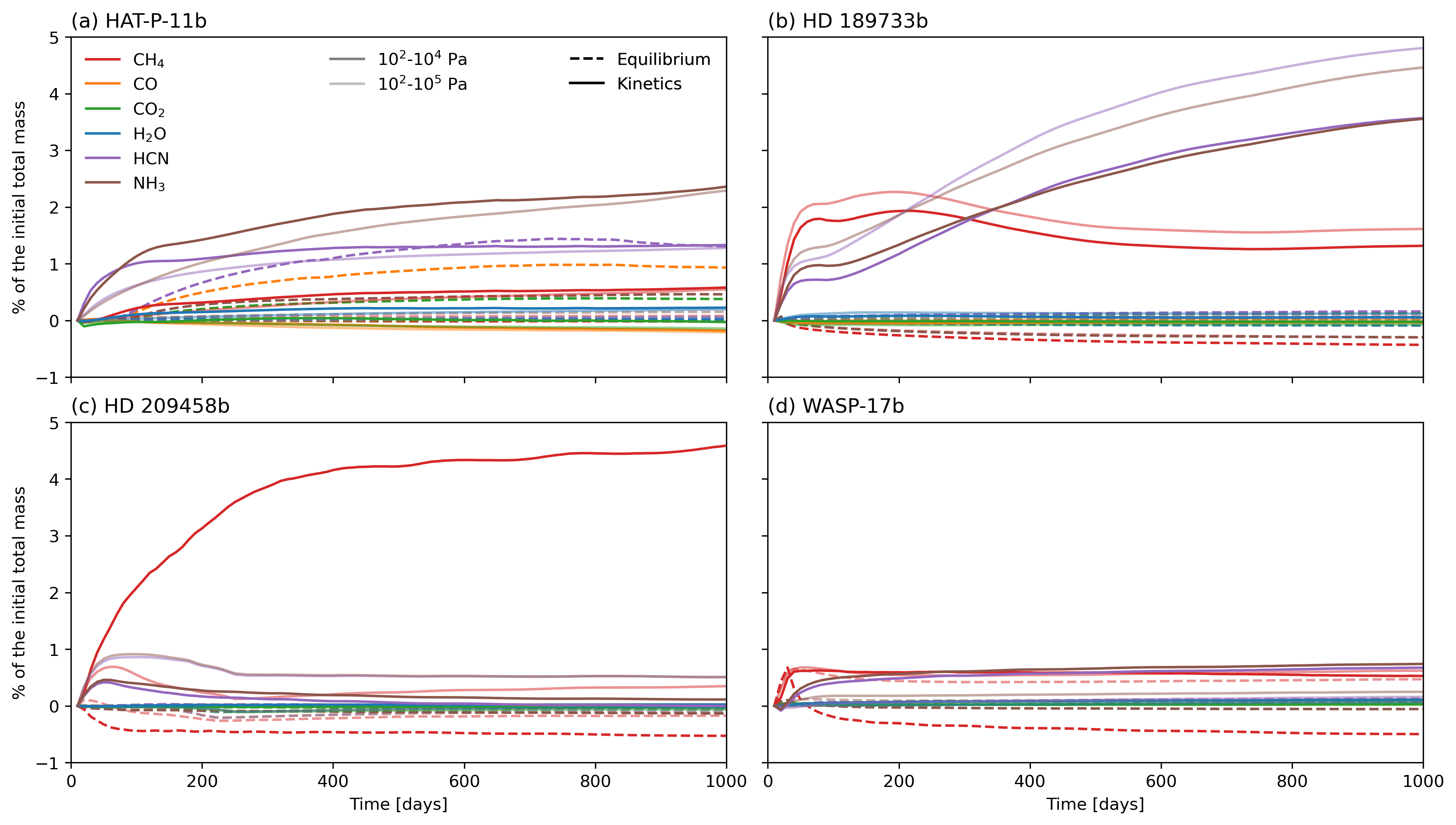}
  \caption{Evolution of the total mass of \ce{CH4}, \ce{CO}, \ce{CO2}, \ce{H2O}, \ce{HCN} and \ce{NH3} within \num{e2}--\num{e4} \si{\pascal} and \num{e2}--\num{e5} \si{\pascal} from the equilibrium (dashed) and kinetics (solid) simulations of HAT-P-11b (a), HD~189733b (b),  HD~209458b (c) and WASP-17b (d).}
  \label{fig:steady_state_ts_burden_pct_of_initial_p1e4_p1e5}
\end{figure*}

To check if the model has reached a steady state, we analysed the evolution of (a) the global maximum wind velocities (Figure \ref{fig:steady_state_ts_wind_max}), (b) the global mean top-of-atmosphere net energy flux (Figure \ref{fig:steady_state_ts_toa_net}), and (c) the total mass of major opacity sources, \ce{CH4}, \ce{CO}, \ce{CO2}, \ce{H2O}, \ce{HCN} and \ce{NH3} (Figure \ref{fig:steady_state_ts_burden_pct_of_initial_p1e4_p1e5}). These quantities serve as metrics of the dynamical, radiative and chemical state of the atmosphere, and no change in these metrics with time indicates that the atmosphere has reached a steady state.

The global maximum zonal, meridional and vertical wind velocities reach 2.6, 1.4 and 0.005 \si{\kilo\meter\per\second}, respectively, for HAT-P-11b, 6.9, 3.5 and 0.01 \si{\kilo\meter\per\second} for HD~189733b, 6.5, 2.7 and 0.02 \si{\kilo\meter\per\second} for HD~209458b, and 6.9, 2.5 and 0.05 \si{\kilo\meter\per\second} for WASP-17b, and change by less than 1\% over the last 200 days. The global mean top-of-atmosphere net energy flux for HAT-P-11b, HD~189733b, HD~209458b and WASP-17b levels off at 0.6, 5, 8 and 23 \si{\kilo\watt\per\square\meter}, respectively, indicating a net inward energy flux equal to less than 5\% of the corresponding incoming energy flux, and changes by less than 1\% over the last 200 days.

The total mass of chemical species evolves differently within each atmospheric layer. It increases if a species is produced within or advected into a layer, and decreases if a species is destroyed within or advected from a layer. Depending on the timescale of processes driving the change in the total species mass, different layers require a different amount of time to reach steady state. To see which layers approach or are at steady state, we integrated the mass of chemical species within two arbitrary chosen pressure ranges (\num{e2}--\num{e4} \si{\pascal} and \num{e2}--\num{e5} \si{\pascal}) and over the entire model domain (\num{e2}--\num{e7} \si{\pascal}). We performed this calculation for both, equilibrium and kinetics, simulations, and as expected, found larger changes in the total species mass in the kinetics simulations discussed in detail below. However, the HAT-P-11b equilibrium simulation was an exception, where the pressure-temperature structure has changed enough on its own to cause the same (1.3\%) increase in the total mass of \ce{HCN} within \num{e2}--\num{e4} \si{\pascal} as in the HAT-P-11b kinetics simulation.

The domain-total mass of chemical species changes little over time because the majority of atmospheric mass is located at high pressure levels (in a regime controlled by chemical equilibrium), making the contribution of the low pressure levels (under potential chemical disequilibrium) to the total relatively small. When deep atmospheric layers are excluded (by integrating mass over a narrower pressure range, e.g. \num{e2}--\num{e5} \si{\pascal}), the total species mass changes more noticeably, especially in the kinetics simulations. For HAT-P-11b, \ce{NH3} and \ce{HCN} increase by 2.3\% and 1.3\%, respectively, from their initial total mass, and for HD~189733b, \ce{CH4}, \ce{NH3} and \ce{HCN} increase by 1.6\%, 4.5\% and 4.8\%, respectively, from their initial total mass. When an even thinner atmospheric layer is considered (\num{e2}--\num{e4} \si{\pascal}), its total species mass evolves differently from its equivalent for a thicker layer, and might reach a different stable mass. For HAT-P-11b, \ce{NH3} and \ce{HCN} increase by a similar amount in this thinner layer relative to a thicker layer. However, for HD~189733b, \ce{CH4}, \ce{NH3} and \ce{HCN} increase by 1.3\%, 3.6\% and 3.6\%, respectively, which is less than their corresponding increases in a thicker layer, and for HD~209458b, \ce{CH4} increases by 4.6\%, which is more than its corresponding increase (less than 1\%) in a thicker layer. These examples illustrate the difference in evolution of the total mass of chemical species in various atmospheric layers. We recommend checking this evolution for atmospheric layers of interest when discussing the proximity of the model simulation to a chemical steady state.

Given the results for the dynamical and radiative metrics mentioned above, and the fact that in all our simulations the total mass of major opacity sources in the atmospheric layer bound by \num{e2}--\num{e5} \si{\pascal} changes by less than 1\% over the last 200 days, we conclude that this atmospheric layer has reached the dynamical, radiative and chemical pseudo-steady state.

\section{Additional figures}
\label{Appendix_C}

\begin{figure*}
\centering
  \includegraphics[width=18cm]{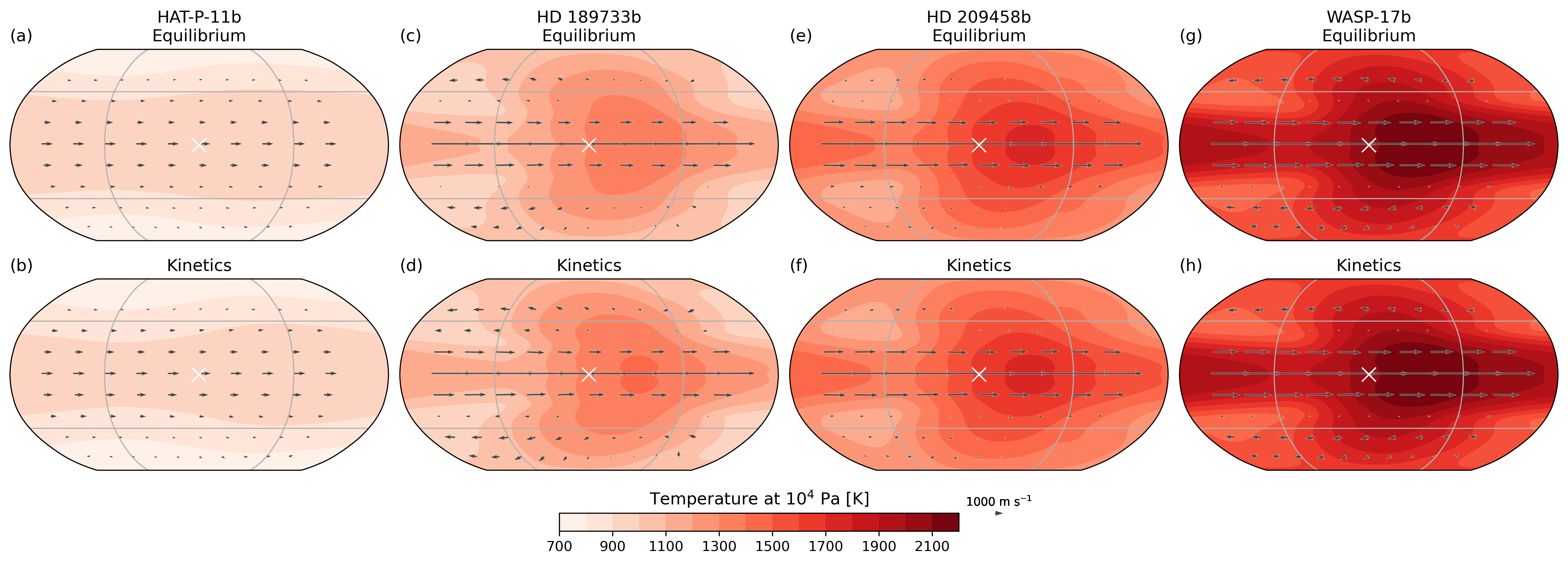}
  \caption{Temperature and horizontal winds at the \SI{e4}{\pascal} pressure level. White cross denotes the substellar point. Grey grid lines show $\pm$\ang{45} latitude and \ang{90} and \ang{270} longitude.}
  \label{fig:temp_hwind_plev_1e4_robinson}
\end{figure*}

\begin{figure*}
\centering
  \includegraphics[width=18cm]{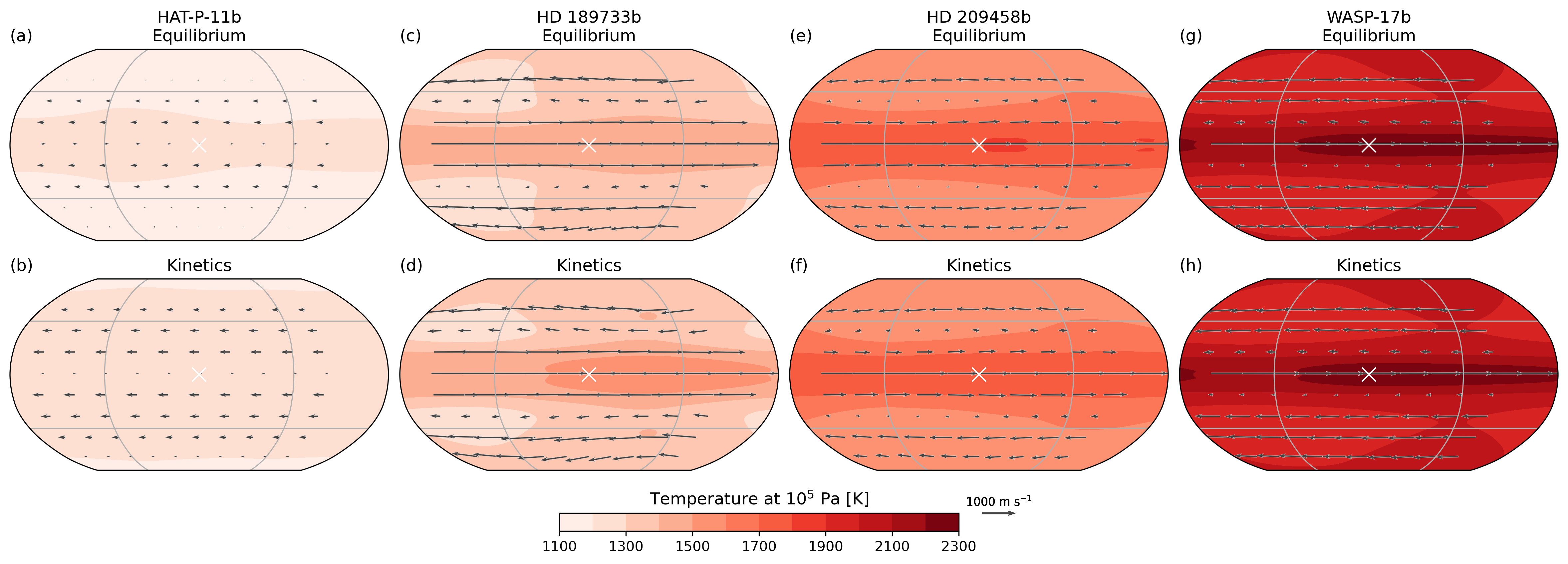}
  \caption{As in Figure \ref{fig:temp_hwind_plev_1e4_robinson} but for the \SI{e5}{\pascal} pressure level.}
  \label{fig:temp_hwind_plev_1e5_robinson}
\end{figure*}

\begin{figure*}
\centering
  \includegraphics[width=17cm]{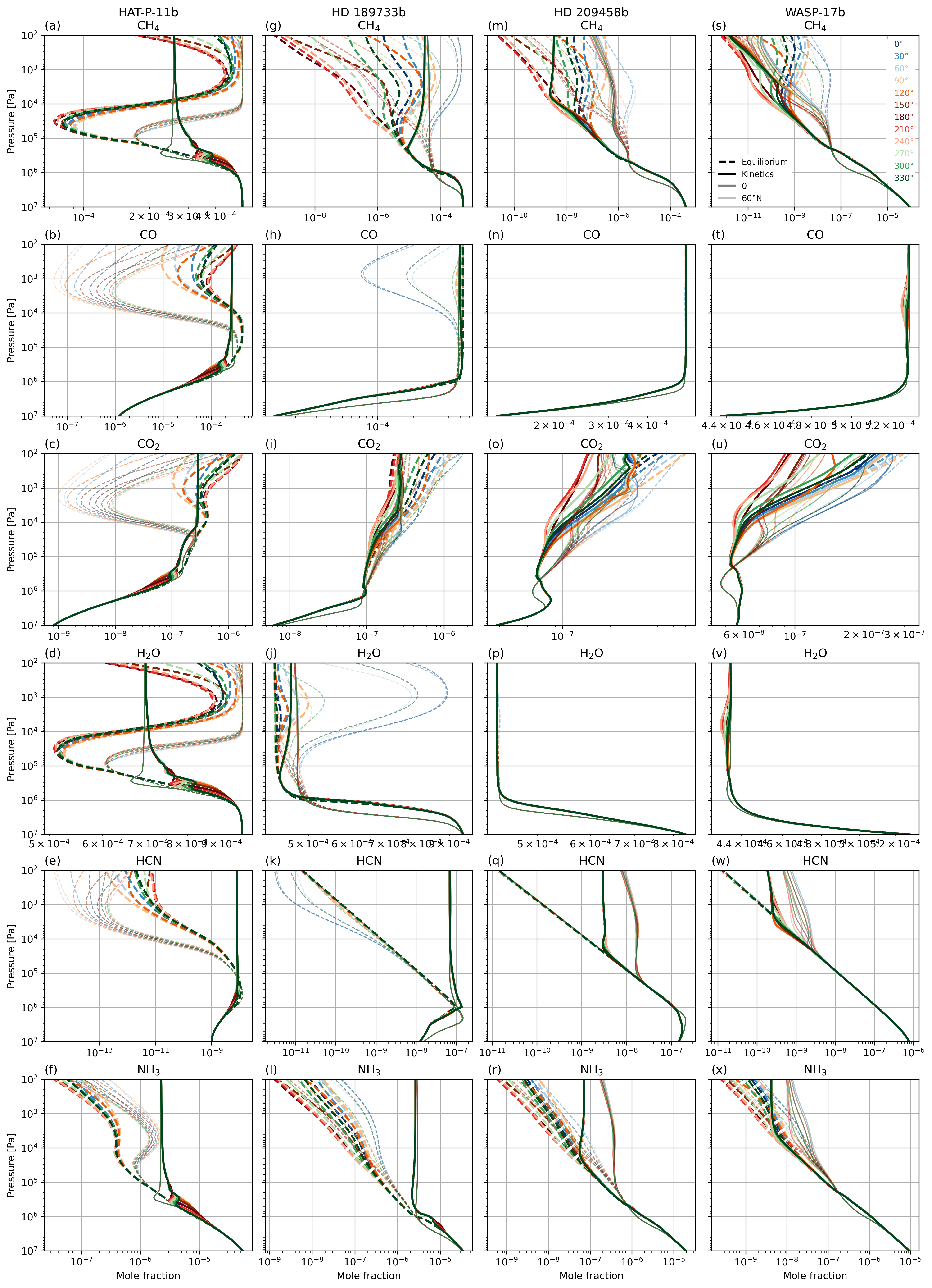}
  \caption{As in Figure \ref{fig:pres_chem_vp_xaxis_same} but with different X-axes.}
  \label{fig:pres_chem_vp_xaxis_diff}
\end{figure*}

\begin{figure*}
\centering
  \includegraphics[width=17cm]{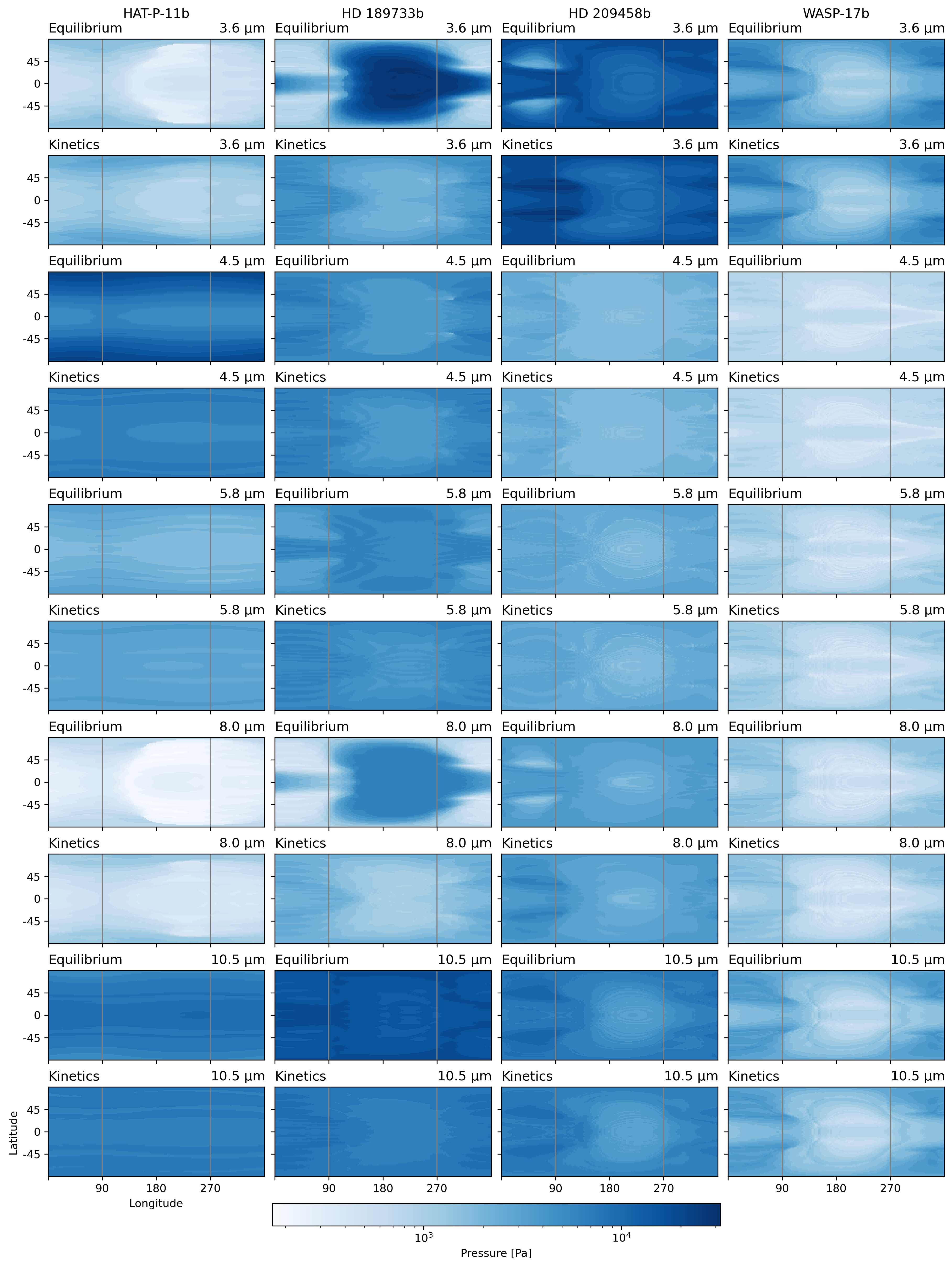}
  \caption{Pressure at a pressure level of the photosphere in individual spectral bands closest to 3.6, 4.5, 5.8, 8.0 and 10.5 \si{\micro\metre} associated with the peaks in absorption by \ce{CH4}, \ce{CO}, \ce{H2O}, \ce{CH4} and \ce{NH3}, respectively, for the equilibrium and kinetics simulations of HAT-P-11b (first column), HD~189733b (second column),  HD~209458b (third column) and WASP-17b (fourth column). The substellar point is located at \ang{180} latitude. Grey grid lines show \ang{90} and \ang{270} longitude.}
  \label{fig:pres_at_lw_norm_flux_con_func_of_unity}
\end{figure*}

\begin{figure*}
\centering
  \includegraphics[width=17cm]{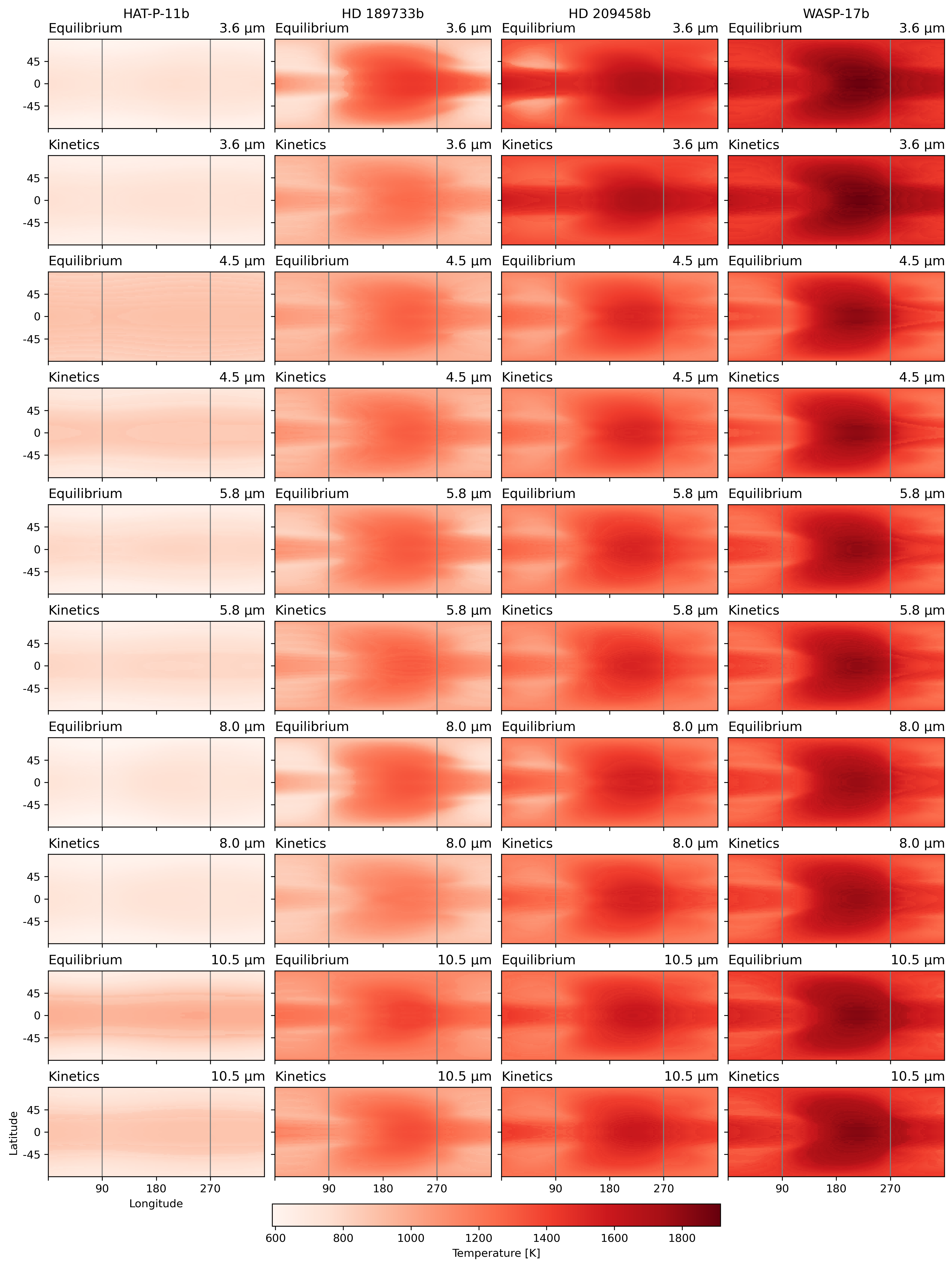}
  \caption{As in Figure \ref{fig:pres_at_lw_norm_flux_con_func_of_unity} but for temperature.}
  \label{fig:temp_at_lw_norm_flux_con_func_of_unity}
\end{figure*}

\begin{figure*}
\centering
  \includegraphics[width=18cm]{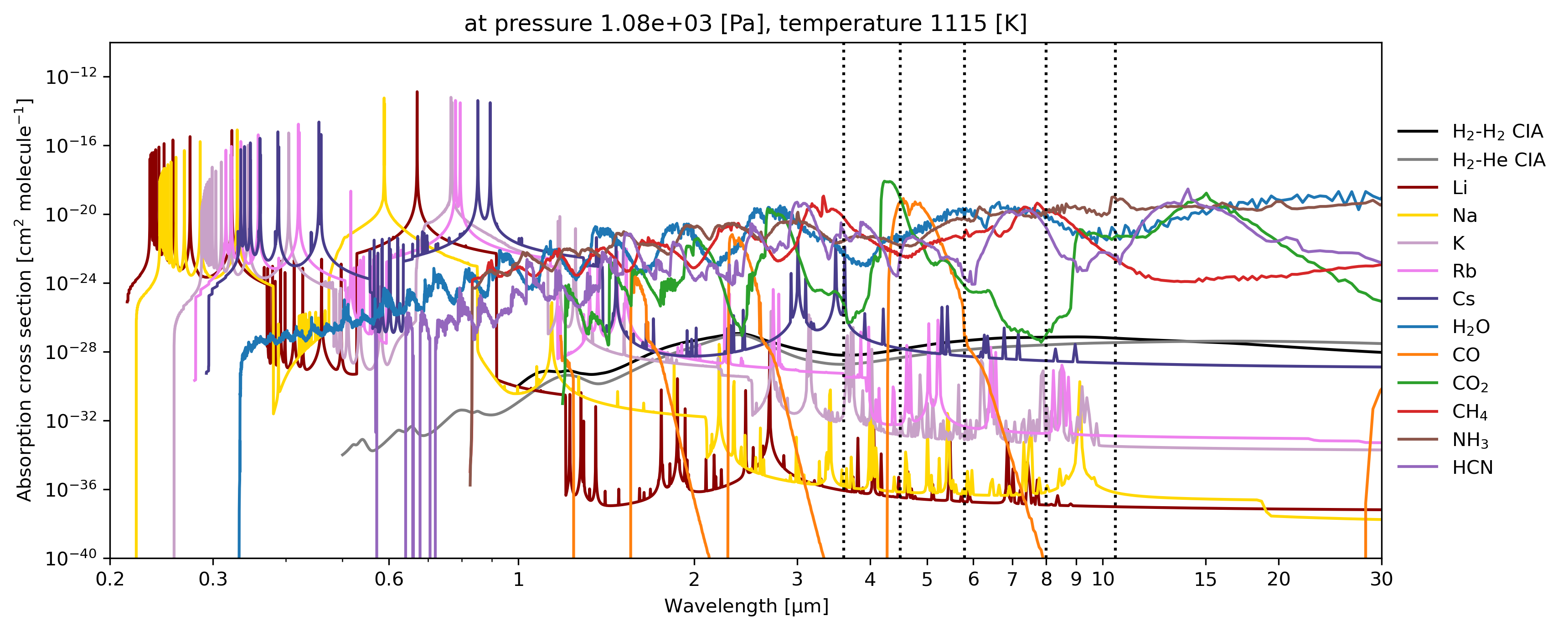}
  \caption{Absolute absorption cross sections of the opacity sources included into our UM simulations in this study (\ce{H2}-\ce{H2} CIA, \ce{H2}-\ce{He} CIA, \ce{Li}, \ce{Na}, \ce{K}, \ce{Rb}, \ce{Cs}, \ce{H2O}, \ce{CO}, \ce{CO2}, \ce{CH4}, \ce{NH3} and \ce{HCN}) at \SI{1.08e3}{\pascal} and \SI{1115}{\kelvin} calculated from the SOCRATES 5000 band spectral files following the methodology of \citet{Goyal2020}. Black dotted vertical lines highlight wavelengths corresponding to the centres of the Spitzer/IRAC channels and 10.5 \si{\micro\metre}. These absorption cross sections are not exact but are relevant for the 10.5 \si{\micro\metre} photosphere pressures and temperatures in HD~189733b kinetics simulation.}
  \label{fig:absorption_cross_section_relevant_for_hd189733b_kinetics}
\end{figure*}

\begin{figure*}
\centering
  \includegraphics[width=18cm]{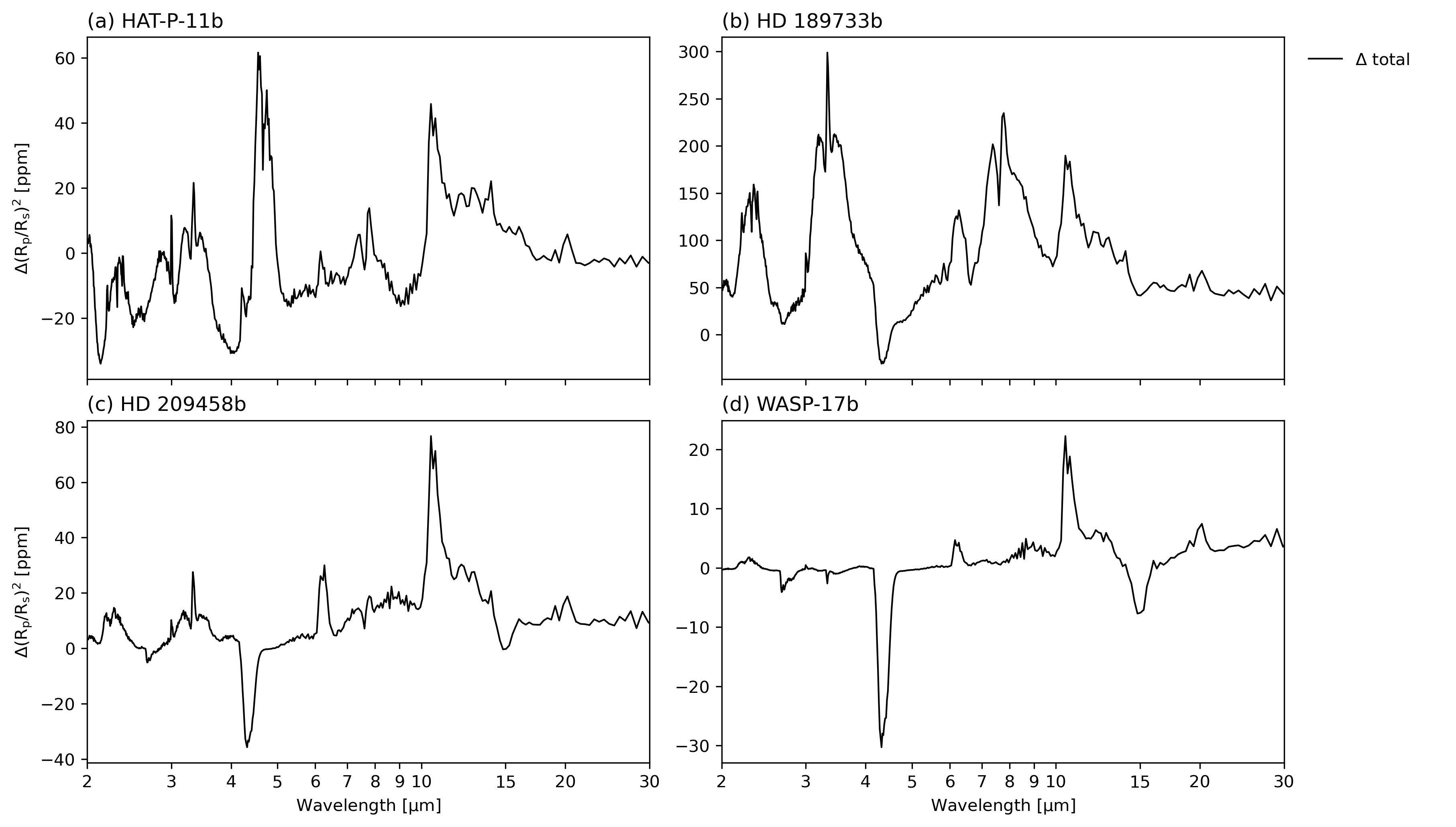}
  \caption{As in Figure \ref{fig:transmission_daynight_total_w_obs} but differences between simulations (kinetics minus equilibrium).}
  \label{fig:transmission_daynight_total_diff}
\end{figure*}

\bsp	
\label{lastpage}
\end{document}